\documentclass{bmcart}

\usepackage{graphicx}
\RequirePackage{hyperref}
\usepackage[utf8]{inputenc} 

\startlocaldefs
\endlocaldefs

\begin{document}

\begin{frontmatter}

\begin{fmbox}
\dochead{Research}


\title{Challenges in Equitable COVID-19 Vaccine Distribution: A Roadmap for Digital Technology Solutions}


\author[
   addressref={aff3,aff10},
   email={joseph.bae@stonybrookuniversity.edu}
]{\fnm{Joseph} \snm{Bae}} \author[
   addressref={aff3},
]{\fnm{Darshan} \snm{Gandhi}} \author[
   addressref={aff3},
]{\fnm{Jil} \snm{Kothari}} \author[
   addressref={aff3},
]{\fnm{Sheshank} \snm{Shankar}} \author[
   addressref={aff3},
]{\fnm{Jonah} \snm{Bae}} \author[
   addressref={aff3,aff11},
]{\fnm{Parth} \snm{Patwa}} \author[
   addressref={aff3},
]{\fnm{Rohan} \snm{Sukumaran}} \author[
   addressref={aff3},
]{\fnm{Aviral} \snm{Chharia}} \author[
   addressref={aff3},
]{\fnm{Sanjay} \snm{Adhikesaven}} \author[
   addressref={aff3},
]{\fnm{Shloak} \snm{Rathod}} \author[
   addressref={aff3,aff13},
]{\fnm{Irene} \snm{Nandutu}} \author[
   addressref={aff3},
]{\fnm{Sethuraman} \snm{TV}} \author[
   addressref={aff9},
]{\fnm{Vanessa} \snm{Yu}} \author[
   addressref={aff3}, 
]{\fnm{Krutika} \snm{Misra}} \author[
   addressref={aff3}, 
]{\fnm{Srinidhi} \snm{Murali}} \author[
   addressref={aff3,aff8},
]{\fnm{Aishwarya} \snm{Saxena}} \author[
   addressref={aff1,aff3,aff5}, 
]{\fnm{Kasia} \snm{Jakimowicz}} \author[
   addressref={aff1,aff2,aff4}, 
]{\fnm{Vivek} \snm{Sharma}} \author[
   addressref={aff3}, 
]{\fnm{Rohan} \snm{Iyer}} \author[
   addressref={aff3},
]{\fnm{Ashley} \snm{Mehra}} \author[
   addressref={aff3,aff6,aff12}, 
]{\fnm{Alex} \snm{Radunsky}} \author[
   addressref={aff3}, 
]{\fnm{Priyanshi} \snm{Katiyar}} \author[
   addressref={aff3},
]{\fnm{Ananthu} \snm{James}} \author[
   addressref={aff3},
]{\fnm{Jyoti} \snm{Dalal}} \author[
   addressref={aff3},
]{\fnm{Sunaina} \snm{Anand}} \author[
   addressref={aff3,aff7}, 
]{\fnm{Shailesh} \snm{Advani}} \author[
   addressref={aff1}, 
]{\fnm{Jagjit} \snm{Dhaliwal}} \author[
   addressref={aff1,aff2,aff3},
   corref={aff1,aff2,aff3},
   email={raskar@mit.edu}
]{\fnm{Ramesh} \snm{Raskar}}


\address[id=aff1]{
  \orgname{Massachusetts Institute of Technology}, 
  \street{02139},                             %
  \city{Cambridge},                           
  \cny{USA}                                   
}
\address[id=aff2]{%
  \orgname{MIT Media Lab, Massachusetts Institute of Technology},
  \street{02139},
  \city{Cambridge},
  \cny{USA}
}
\address[id=aff3]{%
  \orgname{PathCheck Foundation},
  \street{02139},
  \city{Cambridge},
  \cny{USA}
}
\address[id=aff4]{%
  \orgname{Harvard Medical School},
  \street{02115},
  \city{Boston},
  \cny{USA}
}
\address[id=aff5]{%
  \orgname{Ash Center for Democratic Governance and Innovation, Harvard Kennedy School},
  \street{02138},
  \city{Cambridge},
  \cny{USA}
}
\address[id=aff6]{%
  \orgname{Institute for Technology and Global Health},
  \street{02139},
  \city{Cambridge},
  \cny{USA}
}
\address[id=aff7]{%
  \orgname{National Human Genome Research Institute, National Institutes of Health (NIH)},
  \street{20892},
  \city{Bethesda},
  \cny{USA}
}
\address[id=aff8]{%
  \orgname{University of California, Berkeley School of Law},
  \street{94720},
  \city{Berkeley},
  \cny{USA}
}
\address[id=aff9]{%
  \orgname{University of California, San Diego School of Medicine},
  \street{92092},
  \city{San Diego},
  \cny{USA}
}
\address[id=aff10]{%
  \orgname{Renaissance School of Medicine, Stony Brook University},
  \street{11794},
  \city{Stony Brook},
  \cny{USA}
}
\address[id=aff11]{%
  \orgname{University of California Los Angeles},
  \postcode{90095}
  \city{Los Angeles, California},
  \cny{USA}
}
\address[id=aff12]{%
  \orgname{Heidelberg University, Heidelberg Institute of Global Health},
  \city{Heidelberg},
  \cny{Germany}
}
\address[id=aff13]{%
  \orgname{Department of Mathematics, Rhodes University}
  \street{Artillery Rd},
  \postcode{6139},
  \city{Grahamstown},
  \cny{South Africa}
}


\begin{artnotes}
\tiny{
\note{\textsuperscript{1}Massachusetts Institute of Technology, 02139, Cambridge, USA.\\ \textsuperscript{2}MIT Media Lab, Massachusetts Institute of Technology, 02139, Cambridge, USA.\\ \textsuperscript{3}PathCheck Foundation, 02139, Cambridge, USA.\\ \textsuperscript{4} Harvard Medical School, 02115, Boston, USA.\\ \textsuperscript{5}Ash Center for Democratic Governance and Innovation, Harvard Kennedy School, 02138, Cambridge, USA.\\ \textsuperscript{6}Institute for Technology and Global Health, 02139, Cambridge, USA. \\ \textsuperscript{7}National Human Genome Research Institute, National Institutes of Health (NIH), 20892, Bethesda, USA. \\ \textsuperscript{8}University of California, Berkeley School of Law, 94720, Berkeley, USA.\\ \textsuperscript{9}University of California, San Diego School of Medicine, 92092, San Diego, USA. \\ \textsuperscript{10}Renaissance School of Medicine, Stony Brook University, 11794, Stony Brook, USA. \\ \textsuperscript{11}University of California Los Angeles, 90095 Los Angeles, California, USA.\\ \textsuperscript{12}Heidelberg University, Heidelberg Institute of Global Health, Heidelberg, Germany.\\ \textsuperscript{13}Department of Mathematics, Rhodes University, Artillery Rd, Grahamstown 6139, South Africa.} }
\end{artnotes}

\end{fmbox}


\begin{abstractbox}

\begin{abstract} 
The COVID-19 pandemic has led to a need for widespread and rapid vaccine development. As several vaccines have recently been approved for human use or are in different stages of development, governments across the world are preparing comprehensive guidelines for vaccine distribution and monitoring. In this early article, we identify challenges in logistics, health outcomes, user-centric matters, and communication associated with disease-related, individual, societal, economic, and privacy consequences. Primary challenges include difficulty in equitable distribution, vaccine efficacy, duration of immunity, multi-dose adherence, and privacy-focused record keeping to be HIPAA compliant. While many of these challenges have been previously identified and addressed, some have not been acknowledged from a comprehensive view accounting for unprecedented interactions between challenges and specific populations. The logistics of equitable widespread vaccine distribution in disparate populations and countries of various economic, racial, and cultural constitutions must be thoroughly examined and accounted for. We also describe unique challenges regarding the efficacy of vaccines in specialized populations including children, the elderly, and immunocompromised individuals \cite{ho_aamc,crooke2019immunosenescence}. 
Given these complicated issues, the importance of privacy-focused, user-centric systems for vaccine education and incentivization along with clear communication from governments, organizations, and academic institutions is imperative. These challenges are by no means insurmountable, but require careful attention to avoid consequences spanning a range of disease-related, individual, societal, economic, and security domains.
\end{abstract}



\end{abstractbox}
%

\end{frontmatter}



\section{Introduction}
The severity of the COVID-19 pandemic on the health and economies of nations has ushered in an unparalleled period of rapid vaccine research, development, and production. As of April 06, 2022, the WHO reports 153 vaccines undergoing clinical trials and 196 candidates in pre-clinical evaluation \cite{WHOvacc}. When several of these vaccine candidates are approved for widespread use, we anticipate multi-level challenges in their distribution, access, clinical outcomes, patient compliance, privacy-related concerns, and communication. In this work, we divide this landscape of challenges into four distinct categories: 1) logistics, 2) health outcomes, 3) user-centric issues, and 4) communication. We further examine the consequences of these challenges, focusing on the impact for disease spread, individual behavior, society, the economy, and data privacy \cite{hu2021revealing,forman2021covid,jean2021vaccine,aborode2021equal,sharma2020review}. 

The world's manufacturing processes and supply chains are underprepared for the task of widespread vaccine distribution \cite{newton2020covid,alam2021challenges}. Currently, complete frameworks and pipelines for vaccine allocation, distribution, and administration have not been fully developed in all countries internationally, and early proposals fail to account for all potential challenges and obstacles to equitable vaccine coverage \cite{NAP25917,WHOfair,CDCfair}.

Furthermore, there exist many challenges in assessing the effectiveness of a vaccine and its long-term effects on health outcomes. The relative novelty of leading vaccine candidate platforms causes concerns regarding their long-term efficacy and side effects, and the incredible speed of their development must be supplemented by long-term follow-up of vaccinated individuals \cite{clinicaltrials.gov}. There also exist several challenges surrounding user privacy and behavior in the setting of COVID-19 vaccination. Concerns surrounding individual mistrust, the importance of booster vaccination, and data privacy all must be addressed. Finally, several challenges arise in proper communication of vaccine information between governments, companies, and the general public. In the current era of misinformation, mass superstition, and the politicization of science, clear and transparent communication will be vital to the success of widespread vaccination. 

In this early draft, we explore in-depth the various challenges associated with vaccine development and administration in the COVID-19 pandemic. We also provide insight into the consequences of failing to overcome these obstacles, and highlight the need for comprehensive digital solutions. 

\begin{figure}[b]
\centering
\includegraphics[width=0.95\linewidth]{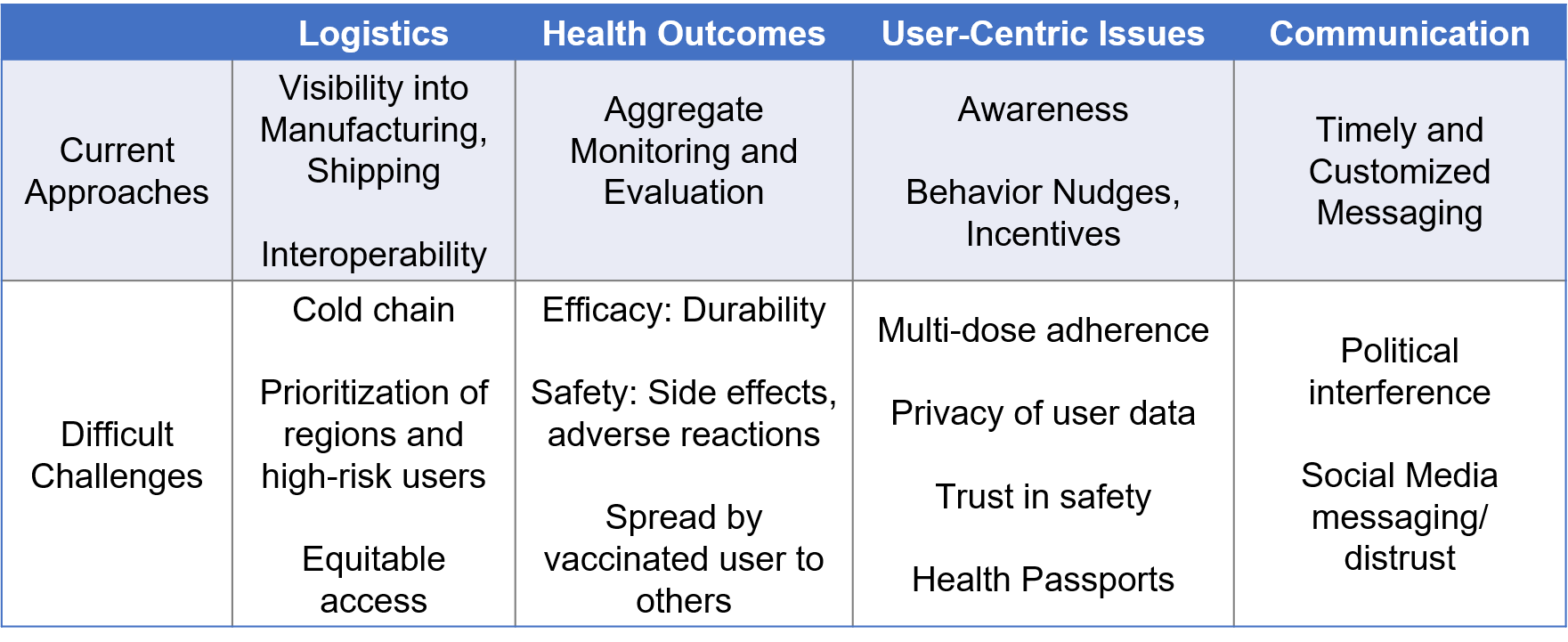}
\caption{Summary of challenges in vaccine distribution.}
\label{fig:SumChall}
\end{figure}

\begin{figure}
\centering
\includegraphics[width=0.95\linewidth]{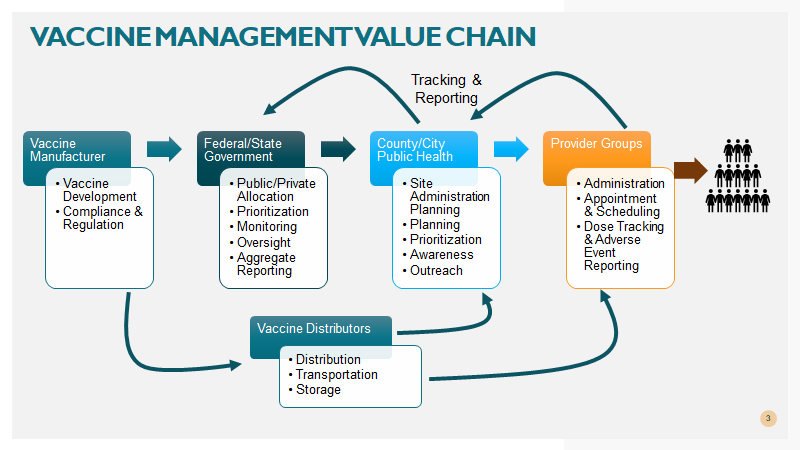}
\caption{A generalized pipeline for vaccine distribution. Challenges can arise at multiple points within this process.}
\label{fig:vaccine_management_value_chain}
\end{figure}

\section{Background}

\subsection*{The Landscape of Vaccine Development}
The current progress in COVID-19 vaccine development can be attributed to both advances in vaccine technologies as well as to historically unparalleled international support and cooperation. In the United States, the Trump administration implemented Operation Warp Speed, a program designed to optimize vaccine regulatory tracts, fund vaccine development, and begin vaccine manufacturing even before approval for distribution \cite{HHS}. International organizations Gavi, the Coalition for Epidemic Preparedness Innovations and the World Health Organization (WHO) have partnered to launch the COVAX initiative, aimed at making vaccines accessible and distributable to all nations \cite{GAVI}. These efforts have contributed to the success of several recent vaccine trial outcomes including Pfizer and Moderna’s recent Phase III trials with reported efficacies of greater than 94\% for each vaccine candidate \cite{modernapress, pfizerpress}. However, challenges still remain in the production, distribution, and monitoring of these and other COVID-19 vaccines. 

\subsection{Challenges identified by a new administration}

President-elect Joe Biden has already established a COVID-19 taskforce that has outlined a general approach for the new administration’s efforts in combating the pandemic \cite{joebidencombatplan, joebidencovid19}. The Biden team references important issues in restoring public trust in science through transparent and educational dialogue, making vaccine availability equitable among different demographics and economic classes, and ensuring transparency in the determination of efficacy and safety of vaccines. However, other challenges remain unaddressed. It is our hope that the present work provides further detail into the many challenges facing widespread vaccine distribution as well as the consequences if those challenges remain unmet. These insights might supplement the future presidential administration's approach to vaccine distribution. 

\subsection{Challenges in vaccination in previous epidemics and pandemics}

The development of vaccines in the setting of previous epidemics and pandemics have encountered numerous challenges including the advent of viral antigenic shifts and antigenic drifts, coordination and communication, and pre-existing health issues. Viral antigenic shifts and drifts occur when a virus strain changes over time, causing a current vaccine to be rendered less effective. Antigenic shifts generally are defined as dramatic, radical genetic changes whereas antigenic drifts occur gradually over time \cite{CDC}. For example, the H1N1 virus underwent an antigenic shift in 2009 enabling it to have much quicker spread when compared to the H1N1 virus observed in 1918 \cite{CDC}. Changes to the virus during and after vaccine development and production can be barriers to effective vaccine coverage of the population at risk. Antigenic shifts can significantly influence vaccine efficacy. \cite{boni_2008} Initial trials of vaccine efficacy can fail to measure vaccine efficacy after important antigenic drift occurs.

Another obstacle for vaccine development is coordination and communication: both on a national and international scale. When dealing with widespread pandemics, it is essential that response forces coordinate together to resolve the pandemic. The importance of coordination was seen in the H1N1 pandemic of 2009, in which communication proved to be both a strong point in the global response and a weak point. Many response forces such as the DOC were successfully activated and international collaboration did occur. However, many of the challenges imposed by the H1N1 pandemic also stemmed from miscommunications \cite{cdch1n1}. For example, the WHO (who was largely responsible for global coordination) was criticized for not taking into consideration research from multiple groups. Due to this, vaccine testing was somewhat crippled, especially in phases 4 and 5 where the WHO convened with only a subset of their committee \cite{fineberg2014pandemic}. There was also ineffective public communication, with a lack of media reports during phase 6 of the testing \cite{fineberg2014pandemic}. While communication can be easily facilitated in countries with large-scale mass media, it is not so easily done in less developed countries with fewer communication resources. In some cases, these countries might rely on direct human communication of medical news and guidelines. The H1N1 pandemic also highlighted the importance of observing infection and vaccination in multiple populations. For instance, it was observed that the H1N1 virus’ effect on pregnant women was quite severe, and safe vaccination of pregnant women was therefore made a priority \cite{vousden2020lessons}. Many of these challenges are also applicable to the future of COVID-19 vaccination development and deployment.

\section{Logistics}

The complexity of planning the production, allocation, distribution, administration, and monitoring of large quantities of a COVID-19 vaccine poses both obvious and non-obvious challenges \cite{GAO}. Of utmost importance is that these challenges are addressed equitably with an understanding of the intersectionality between uniform, large-scale logistical challenges and the varied landscape of governments and populations. We have identified five key logistical challenges that must be addressed for a proper global recovery from the COVID-19 crisis.

\subsection{Challenges}

\subsubsection{Prioritization}

Most frameworks for vaccine distribution suggest that healthcare workers should be the first to receive a potential vaccination, but plans diverge from that point. Organizations including the National Academy of Health, the WHO, and others all have developed their own allocation guidelines \cite{WHOfair,NAP25917,cdc_acip_2020}. A current CDC framework for vaccine distribution does exist, but is limited due to ambiguity in how efficacious novel vaccines might be and how many might be available in initial phases \cite{CDCfair, cdc_acip_2020}. 

Once thorough guidelines are adopted within a given jurisdiction for vaccine distribution, additional challenges may arise in confirming the eligibility of an individual for vaccination. For instance, if an occupational group or individuals with pre-existing conditions are designated for early vaccination, it can be difficult to confirm an individual’s eligibility without overstepping privacy concerns or causing undue logistical burdens. Current medical record systems in the US are non-uniform, making the confirmation of an individual’s pre-existing condition difficult. Furthermore, there are data and privacy concerns that must be considered when distributing vaccines on the basis of health statuses or conditions. This is especially relevant in settings where a patient’s information must be stored in order to facilitate follow-up booster vaccination events or monitoring of long-term side effects.

\subsubsection{Distribution}

There are anticipated difficulties in the pipeline to properly store and distribute a COVID-19 vaccine. The ``cold chain'' references the necessity for some COVID-19 vaccines to be stored at sub-zero temperatures during both transport and shortage \cite{distribution_nys}. Often, dry ice is relied upon in each of these processes, and there is some concern that the US will experience dry ice shortages during the distribution of these vaccines \cite{distribution_nyt}. Pharmaceutical-grade glass vials capable of withstanding sub-zero temperatures might also experience shortages in this setting \cite{distribution_nyt}. In some countries, storage facilities may lack vaccine-qualified refrigerators, potentially greatly affecting the efficacy and storage life of vaccines \cite{distribution_nh}. Furthermore, a proper data framework is not in place to monitor the transport and storage conditions of vaccines distributed widely across multiple countries, and it may be difficult to ensure that a shipment of vaccines has been properly stored throughout transit. This is especially important in the setting of international shipping of vaccines in which storage conditions can be more difficult and costly to maintain. 

In the US, it is more than likely that vaccine transport will rely heavily on shipping companies including UPS and FedEx due to their previous experience in transporting vaccines for illnesses such as the seasonal flu \cite{distribution_nyt}. However, populations that might be most prioritized for vaccination (health-care workers, elderly, those with serious comorbidities, etc.) are not evenly distributed across the US \cite{distribution_wp}. This can pose a challenge in determining where best to allocate vaccine doses and how to properly enact that distribution. An enormous challenge will be the proper coordination of local, state, and federal agencies with manufacturers and shipping companies to ensure proper distribution of vaccines across the country. Improper digital or technological pipelines leave open the opportunity for errors in the proper tracking and transport of vaccine shipments, potentially creating scarcities and surpluses if vaccines are not properly routed to their intended destinations. This is particularly concerning in countries where inconsistencies in internet, satellite, and data storage infrastructure might prevent centralized monitoring of vaccine distribution. 

There also exist issues in determining where and who should be responsible for the vaccination of the general public. In developed nations such as the United States, existing healthcare and pharmaceutical infrastructure will likely be relied upon for dissemination of vaccines. However, issues still remain as to the responsibility of vaccination administration to specific populations such as indigenous populations, federal employees, etc. \cite{distribution_nys}. In developing nations, sufficient levels of centralized vaccination locations for large swathes of the population may not exist. In these cases, an appropriate vaccination framework must be developed. Mass-transit of individuals from remote locations to centralized urban centers for vaccination can have dire consequences for disease spread, but a failure to provide access to vaccines might also marginalize these populations. In these countries, another difficulty may arise in monitoring the mobility of individuals who are constantly migrating. This challenge is compounded in monitoring vaccine distributions in countries where large segments of the population cannot be identified by a standardized identification or address \cite{mugali2017improving,distribution_worldbank}.

\subsubsection{Equity}

Equitable distribution of a potential COVID-19 vaccine is an enormous challenge. This has been highlighted previously in the distribution of the H1N1 vaccine \cite{distribution_la_times}. Economic and racial disparities in vaccine distribution can be exacerbated by inappropriate or unequal eligibility criteria, and it is imperative that government institutions ensure that appropriate guidelines are in place to prevent these disparities. Equitable distribution of the vaccine can also be a challenge on the international scale, as wealthier and more influential nations may have increased access to limited vaccine doses. Prohibitive costs might bring these national differences in vaccine access into sharp focus. Previous data demonstrate discordances in vaccination rates along economic lines both dependent on individual wealth within countries and national wealth on the international stage. Wealthier individuals and nations tend to have higher rates of vaccination whereas less wealthy entities tend to have lower rates \cite{masia2018vaccination,distribution_cdc}. 

\subsubsection{Record Keeping/Follow-up}

Following vaccine administration, obtaining and maintaining secure and thorough records for patients will pose several challenges. These records are crucial in understanding which segments of the population have been vaccinated to guide distribution as well as public health policy. Furthermore, these records might be used to monitor the long-term efficacy and sequelae of vaccination while also ensuring compliance to the two-dose regime currently employed by most vaccine candidate frontrunners. Secure, universal, and accessible health information storage systems are in short supply, and the development of such a system for COVID-19 vaccination recordkeeping will be an enormous challenge. However, proper implementation of such a framework can have enormous societal and individual benefits. 

\subsection{Logistical Consequences}

While the primary effects of unaddressed logistical consequences are plainly ineffective distribution of vaccines, there are nuanced complications surrounding equity affecting multiple spheres of public life. These consequences must be evaluated, and plans must be put into place to alleviate their effects should they become realized. 

\subsubsection{Disease Spread}

A lack of proper digital frameworks to monitor vaccine allocation, distribution, and storage may result in inefficient vaccination of populations, leading to preventable disease transmission. Vaccines that are lost, transported to incorrect geographic locations, or rendered ineffective by improper storage conditions are wasted tools to combat the spread of COVID-19. Furthermore, inadequate monitoring systems can dramatically hinder investigative procedures into vaccine distribution or storage errors, allowing supply chain issues to persist longer than necessary. 

\subsubsection{Individual Behavior}

A few primary consequences might be observed in individual behavior due to inadequate logistical preparation for COVID-19 vaccine distribution. First, individuals in areas where supply chain errors cause reduced vaccine availability will simply be less able to receive a COVID-19 vaccine. These individuals might then be less likely to obtain a vaccine when supply shortages are corrected in their area due to desensitization from the necessity to receive a vaccine. Alternatively, an individual who is unable to obtain a COVID-19 vaccine due to a lack of availability might choose to travel elsewhere for vaccination. This might increase mobility patterns leading to disease spread. Finally, incompetent logistical planning of vaccine distribution might engender mistrust in the government for individuals, potentially decreasing the likelihood of vaccination or compliance with other health policies.   

\subsubsection{Societal Impact}

Consequences noted for individual behavior above might be observed more broadly in societal behaviors, and more widespread mistrust in government competence might be stimulated. Furthermore, logistical inadequacies leading to disparities in vaccine distribution might further widen socio-economic divides between races and classes. 

\subsubsection{Economic Impact}

The economic burden of logistical errors in vaccine production, distribution, and storage can be most obviously understood in the economic cost associated with correcting misdirected vaccines. Considerations for the wasting of vaccines must also be made if transportation or storage conditions are inadequate. Furthermore, these logistical challenges can cause large economic damages due to preventable increases in disease spread and outbreaks, exacerbating the already dramatic economic effects of COVID-19.

\subsubsection{Security/Bad Actors}

Insufficient monitoring of the complete supply chain of vaccine production, distribution, and storage also lends itself to attack by bad actors. If a country does not carefully monitor the total pipeline of vaccine distribution, there may be opportunities for theft and counterfeit vaccine delivery. Furthermore, privacy concerns can arise in numerous areas along the pipeline of vaccine distribution and administration. Private information including health records will undoubtedly be collected for some individuals receiving a COVID-19 vaccine, and solutions must be in place to safeguard against improper usage or access of this data. It is imperative that privacy be safeguarded to the highest possible extent in developing systems to track vaccinated patients and to provide reminders for second-dose administration. 

\subsection{Need for Technology Solutions}

It is imperative that systems are put in place to coordinate and monitor the distribution and administration of COVID-19 vaccines. Vaccine distribution must be coordinated across a variety of private and public sectors including governments, manufacturers, and transport agencies. Constant tracking and monitoring of where vaccine shipments are currently located both in space and in the distribution pipeline must be achieved through a comprehensive, uniform digital framework. Furthermore, systems must be put in place to identify errors in vaccine storage. Additionally, steps must be taken to ensure equitable distribution of vaccines. Ideally, this will be achieved using a transparent methodology that enables the public to directly observe that there are no unintended biases in vaccine distribution. Finally, a privacy-focused approach to vaccination and patient follow-up must be implemented to observe the efficacy of vaccination and to develop understandings of vaccination rates and follow-up vaccination adherence. This record-keeping system will also be crucial in understanding the long-term efficacy and effects of vaccines in diverse populations.

It is also important to note that poorer nations suffer more acutely from many of the challenges outlined in vaccine distribution and administration, indicating the great importance of the deployable digital solutions outlined above.

\section{Health Outcome Challenges}

There is a great deal that is yet unknown about COVID-19 and the vaccines being developed to contain it.  With time, more empirical clinical data will show the long-term efficacy of these vaccines.  Similar vaccines and viruses are also use cases to better anticipate and understand the challenges these newly developed vaccines may encounter. 

\subsection{Novel Vaccine Technologies}

The COVID-19 pandemic has inspired the use of novel vaccine platforms. These technologies have enabled the observed rapid development of promising vaccine candidates, but the fact they have not been widely used previously requires a degree of caution and study as they are implemented. Key concerns are the potential reactogenicity of mRNA vaccines, as well as potential issues in multiple-dose regimens and HIV susceptibility for adenoviral vector vaccines. \cite{fauci2014immune,white_could_2020,buchbinder2020use}.

\subsubsection{mRNA Vaccines}

mRNA vaccines are a very new technology that has been suggested to be ideal for rapid vaccine development in the setting of pandemics. To date, there have been no mRNA vaccines authorized for widespread public use in preventing any viral illness, but previous clinical trials have examined their efficacy with promising results \cite{rauch2018new}. 

Potential challenges previously observed in the setting of other mRNA vaccines include serious reactogenic effects to vaccination as well as potential concerns over relatively weaker immunogenic responses using this platform \cite{pardi2018mrna}. mRNA vaccines for COVID-19 that are developed will need to be observed closely to determine whether reactogenic effects are more prevalent in certain populations, and whether the length of conferred immunity is substantial enough to be useful for public health applications. 

\subsubsection{Adenoviral Vector Vaccines}

Another relatively new technology for COVID-19 vaccine development involves the usage of adenovirus vectors containing genetic coding sequences for COVID-19 antigens. While a more extensively studied platform than mRNA vaccines, there are very few vaccines on the global stage that have been approved for use in any illness using adenoviral vectors. In the setting of COVID-19, Johnson \& Johnson, as well as Astrazeneca, are currently developing vaccines using various adenovirus platforms; both candidates have recently resumed trials following adverse events that were deemed to be most probably unrelated to vaccination \cite{ho_j_and_j,ho_astrazeneca}. 

The primary challenge to adenoviral vector vaccines lies in potential prior host immunity to the viral vector, potentially reducing the efficacy of the original or booster vaccinations \cite{thacker2009strategies}. Furthermore, prior host immunity to adenoviral vectors is nonuniform; for example, some studies have suggested a larger serological immunization prevalence in sub-Saharan Africa for several adenovirus subtypes when compared to Europe and the United States \cite{ho_cen}. If trials do not adequately account for this fact, an adenovirus vaccine might be dramatically less effective in these populations. 

Additionally, two previous adenoviral vector vaccines for HIV were found in clinical trials to increase susceptibility to HIV infection in certain vaccinated patients \cite{fauci2014immune}. There is some concern that the use of adenoviral vectors for COVID-19 vaccines might induce this same susceptibility in some demographics \cite{buchbinder2020use}. 

Finally, there is some indication that viral vector vaccines may be contraindicated in patients receiving certain drug regimes such as some used in the treatment of multiple sclerosis \cite{zheng2020multiple}. The effects of drugs on vaccinations are not currently thoroughly studied but should be considered, especially in the setting of a relatively nascent technology such as adenoviral vectors. 

\subsubsection{Duration of Immunity} 

The duration for which a new COVID-19 vaccine might confer immunity is currently unknown. While clinical trials have thus far demonstrated the safety and efficacy of vaccine candidates, there is still a lack of understanding of how long any vaccine might produce immunity to COVID-19. Currently, estimates are best made based upon the immune response to other, similar coronaviruses as well as studies on the lasting serological response to natural COVID-19 infection \cite{edridge2020seasonal,dan2020immunological}. Each of these suggests that immunity to COVID-19 may not last more than 12 months, and may be even shorter in vaccinated individuals \cite{cruz2021duration,ortega2021seven,edridge2020seasonal,dan2020immunological,wendelboe2005duration}. If the duration of immunity is not well-understood prior to widespread vaccine distribution, it will be imperative for proper tools to be in place for the monitoring of new COVID-19 infection in vaccinated individuals. In scenarios where vaccines for COVID-19 only confer short-lived immunity, protocols to most ideally vaccinate the population at regular intervals will need to be developed and supported by appropriate tools.

\subsubsection{Efficacy of Immunity}

Though current trials have reported high efficacies of greater than 90\% for the Pfizer and Moderna COVID-19 candidates, the potential for lower efficacies must be addressed \cite{pfizerpress,modernapress}. It is possible that immunity may wane over time, and previous clinical trials have produced results that were not consistent with the future performance of a drug/vaccine. Furthermore, variation in the efficacy of individual vaccine doses or lots might be observed due to varying production, storage, and transport conditions. In the setting of diverging vaccination efficacy, tools must be in place to monitor individuals across a broad spectrum of demographic and geographic windows in order to identify causes of varied immunity. Should a vaccine be shown to be uniformly less effective than indicated by clinical trials, procedures must be in place for optimal vaccination strategies and coordination with other preventative policies must be established to prevent renewed COVID-19 outbreaks. A vaccine with lower efficacy can still be useful (as observed in the case of influenza vaccines) \cite{sah2018optimizing,okoli2021decline}, but must be coupled with appropriate procedures and policies accounting for this reduced efficacy. 

\subsubsection{Vaccination of Special Populations}

Current COVID-19 vaccine trials have attempted to accurately assess efficacy across sex, age, and race \cite{ho_pfizer}. Thus far, results from several Phase 3 studies appear to be promising \cite{pfizerpress,modernapress}. However, there is still concern over the immediate and long-term efficacy of vaccine immunogenesis in specific populations such as seniors, children, and immunocompromised individuals \cite{ho_aamc,helfand2020exclusion}. Elderly individuals often undergo a process termed ``immunosenescence" which can hinder immune response to vaccination \cite{crooke2019immunosenescence}. While these concerns have prompted companies such as Pfizer to enroll a large number of seniors in their COVID-19 vaccine trial, it is still unknown whether the long-term vaccine efficacy may differ between patients in different age groups \cite{pfizerpress}. In line with FDA guidelines, clinical trials for COVID-19 vaccine candidates in pediatric populations will follow adult trials, and currently no data has been gathered on the use of these vaccines in children under the age of 12 \cite{ho_fda,ho_pfizer}. Individuals may become immunocompromised due to a variety of health conditions, organ transplant operations, or by taking medications for diseases such as cancer. Vaccines for other diseases have had varied efficacies and morbidities in these populations; COVID-19 vaccines remain to be completely studied in this setting. For individuals with some medical conditions, some vaccines may pose a medical risk \cite{NAP25917}. These individuals must not be subjected to penalties for not receiving a vaccine under the ADA \cite{YANG20203184}. 

\subsection{Health Outcome Consequences}

Aside from obvious effects of lower efficacy vaccines or potential side effects, close attention must be paid to the variance of these effects in various demographics and populations. Unequal health outcomes in the setting of vaccination can have dramatic effects across all levels of society and must be prepared for and mitigated. 
\subsubsection{Disease Spread}

The consequences of ineffective or short-lasting vaccines are self-evident. Further, these issues are made more complicated by the potential for varied effects among different populations. It will be crucial to understand whether disparate long-term effects or the length of immunity conferred by vaccines are disparate in different populations (including the special populations outlined above). A failure to identify and understand these disparities could lead to prolonged disease spread in specific communities or populations. 

\subsubsection{Individual Behavior}

Recent polls have suggested that vaccine efficacy and reactogenicity are among the most critical determinants in individual willingness to obtain a COVID-19 vaccine \cite{ho_pew}. Transparency in reporting real-world efficacies and side effects in various populations is imperative in ensuring that individuals are willing to receive vaccines as they become more available. Additionally, the duration of immunity conferred by vaccines must be communicated to individuals to inform behavior. While the CDC and Dr. Anthony Fauci have suggested that mask-wearing and social distancing will still be needed to optimally combat disease spread even after vaccine distribution, it is likely that vaccinated individuals will be laxer in adherence to these guidelines \cite{ho_cnn,ho_cdc_freq}. It is therefore important that the public be kept well-informed of both the efficacy and length of immunity that they should expect to obtain following vaccination.

\subsubsection{Societal Impact}

Differing vaccine efficacy in diverse populations has the potential to create imbalanced prevalences of disease spread and transmission. This might lead to prejudice and social tension among different demographic groups. Varied vaccination efficacy and lengths of immunity might also influence communal adherence to social distancing and mask-wearing guidelines. Finally, misinformation and a lack of transparency regarding the relative efficacy, side effects, and length of immunity for COVID-19 vaccines might engender unnecessary fear, rumors, and concern in populations.

\subsubsection{Economic Impact}

The timeline of immunity and vaccine efficacy must be well-understood to develop cost-efficient public vaccination regimes. Systems must be in place to identify which vaccines are most efficacious in differing populations. Furthermore, liability compensation programs have been instituted by governments to sponsor vaccine development, leaving governments responsible for reparations in the setting of negative health outcomes caused by vaccination. Ultimately, the most powerful effect of vaccine efficacy on the economy lies in the potential of widespread vaccination to return the economy to some semblance of normality. 

\subsection{Need for Technology Solutions}

The rapid pace of COVID-19 vaccine development has been a boon for the fight against disease spread while engendering caution and concern in those worried about the safety and efficacy of a quickly developed product. A uniform system for both top-down and participatory monitoring of health outcomes for vaccinated individuals can overwhelmingly address these concerns. Such a system must prioritize individual privacy while also retaining enough information to disentangle the effects of a COVID-19 vaccine across multiple demographic groups. 

\section{User-Centric Issues}

In addition to high-level issues regarding the logistics of high-volume vaccine distribution and the effectiveness of new vaccine candidates, it is crucial to consider various issues that are directly pertinent to individual behavior. Citizens must choose to become vaccinated, and incentivizing that behavior will require proper solutions to several challenges. 

\subsection{User-Centric Challenges}

\subsubsection{Trust in the System}

According to  a recent study, 71.5\% of individuals surveyed from 19 different countries would be likely to take a COVID-19 vaccine, with 61.4\% relying on their employer’s recommendation \cite{lazarus2020global}. These numbers vary significantly across countries with up to 90\% of those in China and as little as 55\% of those in Russia indicating that they would be likely to receive a COVID-19 vaccine \cite{lazarus2020global}. Governments will need to contend with the challenge of raising these numbers in order to widely distribute vaccines. To do so effectively, there must be an understanding of which demographics, economic classes, education levels, and geographic locations contain individuals least likely to seek vaccination. Gaining user trust relies on a combination of transparency and clear, effective planning from governments. The challenge of creating these clear lines of communication and pandemic response guidelines in a rapidly evolving disease landscape is a challenge that many governments are finding difficulty in overcoming. 

\subsubsection{Follow-up Tracking and Multi-Dose Reminders}

Today, most COVID-19 vaccines require two-dose vaccination schedules \cite{uci_who}. While these multi-dose regimens are certain to require additional costs and logistical challenges, they have been shown to increase immunogenicity in vaccinated individuals \cite{walsh2020safety,graham2020evaluation}. The importance of developing systems to record individuals who have received one or two doses of the vaccine and to remind individuals to receive a second booster vaccine will be of utmost importance. Paper-based reminders introduce a risk of loss of data due to difficulty in managing and recording a physical paper trail. Furthermore, paper-based reminders pose a greater risk for data spoofing and there are confidentiality concerns related to loss of reminders/records. Systems must also address the monitoring and reminding of individuals that are dependent on others for health care and vaccine obtainment including children and disabled individuals. 

\subsubsection{Data Privacy}

The CDC offers various high-level software tools to assist state and local governments including the Vaccine Tracking System (VTrckS) and PANVAX \cite{uci_cdc2,uci_cdc3}. Other digital products have also been developed by Palantir Technologies Inc., Salesforce, the Maryland Department of Health, Accenture, and SAP in order to address various challenges in vaccine distribution and administration \cite{uci_palantir,uci_sap,uci_accenture,uci_maryland}. These tools could be utilized for identifying and allocating vaccines to high-priority populations, but the lack of privacy guidelines in place to protect sensitive patient data is concerning. For instance, several of these companies have neither disclosed what collected private health information might be used for nor whether any privacy safeguards have been set in place to protect an individual’s name, race, location, travel history, and past health records \cite{uci_palantir}. This personal data might be used maliciously in the setting of data spoofing and data breaches.   

\subsubsection{Vaccine Incentivization and Motivation}

While an individual’s own health and well-being is an obvious incentive for both initial and booster-shot vaccination, the government will need to make advances in motivating large portions of the population to become vaccinated. A recent study reported that as many as 49\% of Americans are inclined towards not receiving a COVID-19 vaccine, with primary deterrents being concerns over side effects and low efficacies \cite{ho_pew}. Governments will need to work to motivate these individuals towards vaccination while grappling with ethical concerns regarding individuals' right to refuse vaccination. Incentivization of vaccination could occur through the implementation of both positive and negative reinforcements for those choosing to be vaccinated and those choosing to forgo vaccination, but again the delicate ethical balance between public health and individual freedoms will need to be addressed \cite{uci_bookings}. 

\subsection{User-centric Consequences}

In a world that continues to place increased importance on individual liberties, vaccine frameworks that do not attempt to build user engagement and trust will have dramatic consequences. 

\subsubsection{Disease Spread}

A failure to properly motivate individuals to vaccination and adherence to booster doses can obviously have direct consequences on disease spread. Lower vaccine adoption rates will lead to decreased levels of population immunity and therefore higher disease prevalence. 

\subsubsection{Societal Impact}

Left unaddressed, the above user-centric issues can result in population level trends in mistrust and low rates of vaccination. Vaccination frameworks must be designed with user behavior in mind lest societal norms outweigh scientific and government guidelines. Explicitly, if focused systems are not set in place to guide users through the process of vaccination and follow-up, society will dictate its own acceptable standards for these processes which may not align with official recommendations.  

\subsubsection{Economic Impact}

A balance between resource investment and return must be made in designing an incentive structure for vaccination and adherence to multi-dose vaccination. Additionally, data privacy breaches can result in large-scale fraud and economic loss when instigated by malfeasant parties or result in losses via lawsuits when the result of improper precautionary measures. 

\subsubsection{Security/Bad Actors}

Technologies seeking to address issues in vaccine distribution pipelines must take thorough measures to protect user data and privacy. If proper legislation is not introduced by governments or if companies grow lax in data security measures, there are numerous opportunities for malicious access and use of private health information.

Technologies seeking to address issues in vaccine distribution pipelines must take thorough measures to protect user data and privacy, such as individuals' consent for collecting sensitive information before disclosing their identity. To obfuscate the sensitive personal information, one could guarantee privacy using cryptographic techniques such as hashing, private set intersection, secret sharing and fully homomorphic encryption \cite{enact,epione,Rogue,berke2020assessing,bell2020tracesecure}. The government, public health department, policymakers, or companies should introduce robust data security measures to minimize security risks. Thus, avoiding numerous possibilities for malicious access and use of private health information.

\subsection{Need for Technology Solutions}

Technological solutions and backing are crucial in assisting and protecting against the most important issue of data theft and breach, indirectly helping to create a sense of trust among users. These tools must be personalized, real-time, and come from trusted sources such as public governments. These tools should also enable participation in large observational studies, and enable passive user engagement without unnecessary user input in order to set reminders, schedules, etc. Finally, these tools must be transparent, enabling a user to understand the use of their data. In each of these aspects, digital solutions can be more effective and secure than the use of traditional paper methods for user engagement and follow-up. 

\section{Communication}

Communication is perhaps as important a tool as vaccination itself in combating the COVID-19 pandemic. Fears surrounding a ``COVID theater'' of political communication in which certain public leaders might appear to take stances on issues without substantive action, have resulted in the proliferation of misinformation and uncertainty. The public must be made thoroughly and transparently aware of both the importance of vaccination, as well as the fact that vaccines are only a part of the continued public health framework of social distancing and mask-wearing that must be put in place to overcome COVID-19. 

\subsection{Communication Challenges}

\subsubsection{Political Communication}

Proper governance is integral to a successful reduction in disease prevalence. Political leaders who downplay the pandemic or disregard the opinions of health care experts and scientists in an attempt to manufacture good news to the public can stifle a nation’s trajectory towards recovery \cite{comm_bjm_1,comm_bjm_2}. In the United States, the presidential administration has been criticized for mixed stances on mask-wearing and the use of hydroxychloroquine for COVID-19 treatment \cite{comm_et_healthworld}. These messages have oftentimes conflicted with the stances of the national scientists including Dr. Anthony Fauci, potentially undermining public trust in public health policies \cite{comm_fauci}. A lack of consistent messaging across a country’s top executive and scientific officials creates ideal conditions for the spread of false information and mistrust. 

\subsubsection{Coordination in Vaccine Distribution}

Thorough distribution of a vaccine in both developed and developing nations will require varying levels of coordination between multiple levels of government, private companies, and nonprofit organizations. Within developed countries, a collaboration between national health authorities and authorities at the state, district, and county levels will be required to assess vaccination needs and plan the distribution of vaccines accordingly. Developing countries will often require collaborations with NGOs or organizations such as the WHO, GAVI, etc. to procure and distribute vaccines \cite{comm_who}. In either case, communication between a wide array of governmental, corporate, and nonprofit organizations will be crucial for successful vaccine dissemination. 

\subsubsection{Miscommunication and Trust}

The majority of the American population is divided in the acceptance of vaccines. Both long-standing anti-vaxxer sentiment and newer COVID-19-specific concerns produce reluctance towards use of  COVID-19 vaccine. One widespread concern involves the efficacy and safety of COVID-19 vaccines that appear to have been produced far faster than previous vaccines \cite{comm_stat}. However, it has been demonstrated that the increased speed of vaccine development is most primarily due to improvements in technology as well as the unprecedented levels of funding and government cooperation leveraged for COVID-19 vaccine production  \cite{comm_hastings}. Furthermore, there is larger widespread support for full FDA approval rather than emergency use authorizations (EUAs). Previously, data for EUA decisions were not publicly available, but the FDA has recently agreed to make this data public \cite{comm_cidrap}. These misconceptions have not been well-communicated to the public and form the basis for many rumors and misconceptions regarding vaccine safety.  

Systemic biases and discrimination present in medicine have also impacted communication surrounding COVID-19 vaccine distribution. \cite{comm_stat2} For instance, plans to distribute COVID-19 vaccines to marginalized populations first have sparked fears that such a policy would be viewed as experimentation of the vaccine on minority populations \cite{schmidt2020lawful}. 

\subsubsection{Misinformation Shared on Social Media}

We are facing an `infodemic’ with the surge in rumors and misinformation spread on social media platforms \cite{patwa2020fighting}. Rumors regarding cures for COVID-19 have led to death and hospitalizations~\cite{comm_timesofisrael,patwa2020fighting}, and have also prompted individuals to be lax in mask-wearing and social distancing.  A popular conspiracy theory~\cite{comm_bbc} claims that COVID-19 vaccines may be utilized by Bill Gates to implant individuals with microchips. Obviously, these rumors and theories fuel anti-vaccine sentiment in the population. Social media is the primary breeding ground of such misinformation~\cite{burki2019vaccine}.

\subsection{Communication Consequences}

Unaddressed, challenges in communication threaten the very foundation of widespread vaccination. Miscommunication whether intentional or unintentional has the power to sway individual and societal understandings of vaccine efficacy and safety. Below we outline key effects of miscommunication.

\subsubsection{Disease Spread}

The effect of miscommunication on individual and societal behavior is the primary way in which it will tend to affect disease spread. Miscommunication can reduce vaccination rates, adherence to public health policies, etc., thereby resulting in higher disease prevalence and spread. 

\subsubsection{Individual Behavior}

Improper or inadequate communication by the government can have dramatic implications for an individual's likelihood to adhere to precautionary measures in the pandemic. Misinformation that downplays or underestimates the severity of COVID-19 can de-incentivize people from seeking vaccines. The same effect is produced if vaccine trials and efficacy are not properly and thoroughly explained to the public. Skepticism, doubt, and mistrust can act even more dramatically, causing individuals to actively reject vaccines, further compounding beliefs that the COVID-19 pandemic is ``fake news''. Proper communication and education surrounding the need for two-dose vaccination regimes is also critical in properly providing vaccine coverage to the population.

\subsubsection{Societal Impact}

In America, already inflamed public tensions can become further exacerbated in populations that believe that their own source of information regarding COVID-19 is an absolute fact. In the setting of contradicting or unclear messaging from multiple government and scientific sources, groups might be easily turned against one another as they rely on their own personal source of information. This will engender a positive feedback cycle of mistrust and further antagonism between groups at odds with one another. An illustrative example lies in the use of masks; even some physicians in the US have described their belief in the inefficacy of masks, furnishing groups in society with backing for anti-masker views \cite{comm_bbc2}.

\subsubsection{Economic Impact}

Miscommunication can cause multiple sources of economic loss. Small businesses can be affected by miscommunicated policy guidelines and difficulty enforcing policies that are not believed by the public to be necessary, as in the case of wearing masks in some areas. Miscommunication can also result in an improper allocation of resources or goods; for instance the advertisement of false COVID-19 therapies can result in wasted resources in obtaining ineffective treatments.

\subsubsection{Security/Bad Actors}

Miscommunication and the spread of false information is one of the primary ways in which bad actors can influence behavior in the setting of the pandemic. If governments are not explicit about vaccination procedures, it is possible for bad actors to provide disingenuous information in order to perform fraud or other malicious actions. For instance, if long-term monitoring is introduced for vaccinated individuals, false apps or websites as were observed in the setting of contact tracing might appear in attempts to steal user information \cite{comm_tradingstandards}. These same security breaches can occur at every intersection of user interaction with government protocol if guidelines are not explicitly and transparently communicated to individuals. 

\subsection{Need for Technology Solutions}

Technological advances have the potential to greatly augment effective communication efforts during COVID-19 vaccine distribution. Large social network companies are in positions to take bigger steps to eliminate the spread of misleading information (that could potentially discourage vaccine adoption) through their platforms. Social media platforms also are uniquely situated to disseminate accurate educational information concerning vaccine development and distribution. Furthermore, advances in machine learning-based fake news detection \cite{patwa2020fighting, Vijjali_transformer_covid_fake_news} can help automate the process of removing misinformation. Finally, it is critical that governments have intense security built around their technological solutions to prevent unwanted influence from bad actors.
	
\begin{figure}
\centering
\includegraphics[width=0.95\linewidth]{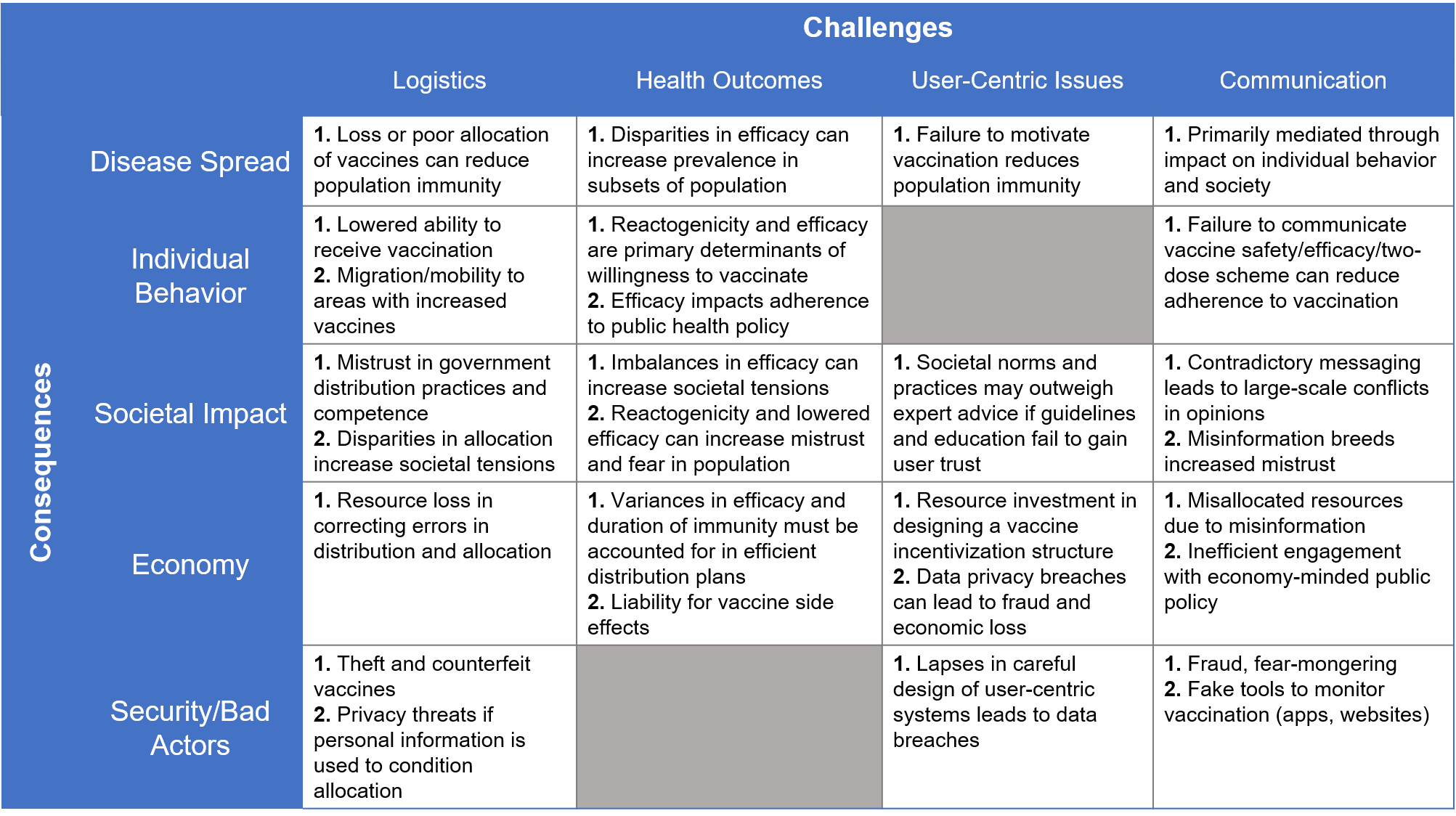}
\caption{Summary of challenges and consequences of vaccine distribution}
\label{fig:challenge_consequence_matrix}
\end{figure}

\section{Challenges in Various Countries}

While the H1N1 pandemic was erupting in the spring of 2009, the United States and other rich countries paid to reserve \cite{brown_2009} virtually the entire supply of potential vaccines — agreeing to release 10\% of it to developing countries only after it became clear months later that the virus was less deadly than originally thought. Today, the planet seems just as unlikely to operate cooperatively. When the COVID-19 pandemic struck, more than 70 countries seized control of medical supplies to prevent them from being shared or sold abroad \cite{world_trade_oranization_2020}. After the antiviral drug Remdesivir showed promise, the United States bought 90\% of the global supply. Already, wealthy countries that makeup just 13\% of the global population have pre-purchased more than half of the vaccine that major companies have pledged to produce, according to the aid organization Oxfam. The United States pre-purchased 800 million doses of various experimental vaccines in a bet that at least one candidate will prove safe and effective.

\subsection{Middle East}

The Middle East has been greatly affected by the presence of ongoing conflict, leading to widespread inequalities in healthcare provisions \cite{fawcett2021middle}. Many Middle Eastern countries have low vaccination rates due to higher vaccine hesitancy, limited vaccine deliveries from the COVAX facility, and a relative difficulty of purchasing vaccines during a period of economic decline which results due to the pandemic \cite{Delta_2021}. Furthermore, ongoing conflicts have made many countries hard to reach, hampering vaccine delivery to the region. For example, only 2\% of Yemen's population have received at least a single dose, with Syria having around 5\% of people receiving one or more doses \cite{Middle_East_Yemen}. This has coincided with the ongoing civil conflict within both Yemen and Syria. Similar effects are observed within Afghanistan, as vaccine supply has been limited. Although Afghanistan was expected to receive more doses, the current insurgency has limited the supply of vaccines to Afghanistan \cite{essar2021covid}. As such, only around 15\% of the population in Afghanistan have received one or more doses \cite{Afghanistan_WHO_Portal}. In Iraq, vaccine hesitancy has proven to be a barrier to vaccine access. There has been a lack of governmental trust and widespread misinformation about the vaccine that has contributed to a low vaccination rate in Iraq (22\%) \cite{COVID_Iraq_2021}.

\subsection{Africa}

Most African countries are often back in the queue regarding new technologies and public health interventions, including vaccines. Low COVID-19 vaccination coverage in Africa owes to vaccine nationalism, diplomacy and vaccine hesitancy. 
\subsubsection{Equitable Vaccine Distribution Challenge: Inability of COVAX meeting targets}

Currently, there are many challenges to vaccination in African Countries, due to which most African countries missed WHO’s December 2021 target of achieving full vaccination rates of 40\% in every country by a wide margin. Only 9\% of Africa were fully vaccinated against Covid by 2021 \cite{unicef_2021, COVID_vaccination_2021}. A challenge to vaccination in African countries as seen by most countries was the inability of COVAX initiative to receive the estimated number of doses for 2021. This was partly connected to the global community's hypocrisy. Most countries depend on importing vaccines from outside the continent, mainly from India and the US. Wealthier countries inked arrangements with vaccine producers as early as July 2020, while the vaccines were still being developed and tested. Manufacturers gave them top priority, making it impossible for the COVAX programme, the African Union, and individual nations to get dosages. Despite the fact that the international community recognises the need for worldwide vaccination, there is no strong commitment to hasten vaccine deployment to Africa \cite{karafillakis2017benefit}.

The WHO and the African Centers for Disease Control and Prevention (CDC) have reported that the timing at which the COVAX dosages arrive is unpredictable. Instead of a more consistent, continuous supply, African countries have witnessed a ``huge dump'' of dosages many times during the year. Despite the fact that advocates for vaccination equity encourage high-income nations to contribute even more vaccines, few countries provided vaccines with a shelf life shorter than the WHO's two-and-a-half-month limit. 450,000 expired doses were discarded because a handful of African countries were unable to administer them quickly enough.

\subsubsection{Challenges of Poor Health Infrastructure, Lack of funds for Personnel training and Vaccine Storage}

However, the major reasons for low vaccination in African countries were poor health infrastructure, a lack of funding for medical personnel training and deployment \cite{who_2021}, and vaccination storage problems \cite{mwai_2021}. A crucial problem for getting vaccines to LMICs, particularly in Africa, is an interrupted cold chain. Here, most communities live without continuous power supplies, and freezers that cost up to \$20,000 are unaffordable \cite{equity2021covid}. According to the WHO, out of the total vaccines supplied in Africa, only 63\% have been administered (by the end of 2021). Moreover, more than half of the African countries have used fewer than 50\% doses that were supplied. Following the discovery of the Omicron variant in South Africa and its fast worldwide spread in recent weeks, these low vaccination rates have been of particular concern.

\subsubsection{Reluctance/ Hesitancy to Vaccines}

Low vaccination coverage and widespread vaccine hesitancy among a significant fraction of the population in Africa, in particular, jeopardize attempts to combat the COVID-19 pandemic. Trust in contemporary vaccinations has been eroded by a history of colonial medical and vaccine research abuse in Africa \cite{mutombo2021covid}. Misinformation and a lack of nuanced and culturally aware understandings of vaccination reluctance are also key contributors \cite{sallam2021covid}. This is particularly dominant in African health professionals which accounted for low vaccination rates among them. The same was confirmed by the WHO in November 2021. 

Many nations on the continent including Nigeria (5.8\%), Sahrawi republic (0.1\%), Democratic Republic of Congo (0.6\%) have only vaccinated less than 10\% of their people \cite{COVID_vaccination_2021}. Due to worries about blood clots, the DR Congo has postponed the introduction of AstraZeneca vaccinations. The government then returned 13 million doses to the COVAX Facility after commencing its campaign since they could not be delivered before they expired. South Sudan too returned most dosages to the facility before they expired \cite{jerving2021long}. The WHO has formed a goal of 70\% coverage for all countries by June 2022, although this seems to be difficult for most African countries. Some countries resisted launching a vaccination campaign. Tanzania's government began its campaign at the end of July, signaling a shift in policy. Burundi started vaccination very late on 18th Oct, 2021. Eritrea has not yet begun a vaccination programme \cite{Reuters_2021}.

\subsubsection{Low Clinical Trial in Africa (<2\%) Raising Efficacy Issues}

Another serious concern that has yet not been highlighted much is the low number of clinical trials done in Africa. This may have implications when it comes to efficacy studies of vaccines for populations across the continent. Out of the clinical vaccine trials, less than 2\% have been conducted in Africa. The same was put forward with concern by the African Academy of Sciences \cite{makoni2020covid}.

\section{COVID-19 Vaccination Booster Dose}

Following the deployment of mass vaccination efforts, the scientific community and policymakers have been confronted with new challenges, notably regarding the possible fading of vaccine efficacy. It is uncertain if supplementary doses are required, and researchers are seeking to understand whether, when, and for whom booster doses might be beneficial for preventing COVID-19 and pandemic control \cite{croda2021booster}. However, recent vaccine clinical trials have shown that booster doses may be required even in the general population due to waning immunity caused by the primary vaccination. A secondary reason is that variants expressing new antigens have evolved to the point where immune responses to the original vaccine antigens no longer adequately protect against currently circulating viruses \cite{krause2021considerations}.

Israel first rolled out Pfizer-BioNTech vaccine booster doses for at-risk populations in July 2021 \cite{Lieber_2021}. In Israel, vaccinated persons in January or April had lesser effectiveness against severe sickness than those immunized in February or March. A short-term protective benefit of a third dosage (compared to two doses earlier) was observed during the first three weeks of August 2021, right after booster doses were widely implemented \cite{krause2021considerations}. In the US, the CDC began administering booster doses to immuno-compromised persons in summer 2021 \cite{CDC_2022}. Vaccine equity experts have advocated that since primary vaccination programs in most countries are still in preliminary stages, the available vaccine quantities should first be used to vaccinate high-risk persons with severe illness that have not yet been immunized. Even if any advantage may be derived via boosting in the long run, it will not exceed the benefits of giving early protection to the unvaccinated, thus saving more lives \cite{krause2021considerations}. WHO has asked for a halt to booster doses until the advantages of primary immunization are made available to more individuals worldwide \cite{WHO_Director-General_2021}.

\section{Discussion and Conclusion}

Here we present four broad categories of challenges facing vaccine distribution and monitoring in the COVID-19 pandemic in addition to five consequences that these challenges might cause. Our work is intended to succinctly compile these findings for governments, organizations, and individuals as they navigate the unprecedented landscape of mass vaccine distribution. This is not intended to be an exhaustive enumeration of every challenge likely to be faced in this setting, but we hope to address those challenges with the most potential to negatively impact society should they remain unaddressed. Future work will seek to better identify the solutions and frameworks already developed to meet these challenges, in addition to proposing privacy minded digital solutions to many of the obstacles demonstrated in this work.
	
The initial step in overcoming the many challenges associated with vaccination in the COVID-19 pandemic is the proper identification of each potential issue. In this early draft, we have presented challenges spanning the logistics, health outcomes, user-centric issues, and communication surrounding COVID-19 vaccine development and distribution. We explicitly connect these challenges to consequences in disease spread, individual behavior, societal impact, economic impact, and data security. We hope that the insights illustrated in this paper provide both policymakers and individual patients further information into the many challenges associated with vaccination for COVID-19.

Many challenges remain unaddressed by current vaccination strategies and guidelines, and, in some cases, no easy solutions exist. Nevertheless, it is crucial that governments, organizations, academics, and individuals begin to develop the frameworks and approaches necessary to prevail over these challenges. Advances in technology and global cooperation have brought about the development of incredibly efficacious vaccine candidates. These same forces must be harnessed in a transparent, privacy-minded, user-centric manner in order to ensure the equitable and effective distribution of COVID-19 vaccines to all.


\begin{backmatter}

\section*{Competing interests}
  The authors declare that they have no competing interests.


\section*{Acknowledgements}
We are grateful to Riyanka Roy Choudhury, CodeX Fellow, Stanford University, Adam Berrey, CEO of PathCheck Foundation, Dr. Brooke Struck, Research Director at The Decision Lab, Canada, Vinay Gidwaney, Entrepreneur and Advisor, PathCheck Foundation, and Paola Heudebert, co-founder of Blockchain for Human Rights for their assistance in discussions, support and guidance in writing of this paper.


\bibliographystyle{bmc-mathphys} 
\bibliography{bmc_article}      


\begin{thebibliography}{117}
\ifx \bisbn   \undefined \def \bisbn  #1{ISBN #1}\fi
\ifx \binits  \undefined \def \binits#1{#1}\fi
\ifx \bauthor  \undefined \def \bauthor#1{#1}\fi
\ifx \batitle  \undefined \def \batitle#1{#1}\fi
\ifx \bjtitle  \undefined \def \bjtitle#1{#1}\fi
\ifx \bvolume  \undefined \def \bvolume#1{\textbf{#1}}\fi
\ifx \byear  \undefined \def \byear#1{#1}\fi
\ifx \bissue  \undefined \def \bissue#1{#1}\fi
\ifx \bfpage  \undefined \def \bfpage#1{#1}\fi
\ifx \blpage  \undefined \def \blpage #1{#1}\fi
\ifx \burl  \undefined \def \burl#1{\textsf{#1}}\fi
\ifx \doiurl  \undefined \def \doiurl#1{\textsf{#1}}\fi
\ifx \betal  \undefined \def \betal{\textit{et al.}}\fi
\ifx \binstitute  \undefined \def \binstitute#1{#1}\fi
\ifx \binstitutionaled  \undefined \def \binstitutionaled#1{#1}\fi
\ifx \bctitle  \undefined \def \bctitle#1{#1}\fi
\ifx \beditor  \undefined \def \beditor#1{#1}\fi
\ifx \bpublisher  \undefined \def \bpublisher#1{#1}\fi
\ifx \bbtitle  \undefined \def \bbtitle#1{#1}\fi
\ifx \bedition  \undefined \def \bedition#1{#1}\fi
\ifx \bseriesno  \undefined \def \bseriesno#1{#1}\fi
\ifx \blocation  \undefined \def \blocation#1{#1}\fi
\ifx \bsertitle  \undefined \def \bsertitle#1{#1}\fi
\ifx \bsnm \undefined \def \bsnm#1{#1}\fi
\ifx \bsuffix \undefined \def \bsuffix#1{#1}\fi
\ifx \bparticle \undefined \def \bparticle#1{#1}\fi
\ifx \barticle \undefined \def \barticle#1{#1}\fi
\ifx \bconfdate \undefined \def \bconfdate #1{#1}\fi
\ifx \botherref \undefined \def \botherref #1{#1}\fi
\ifx \url \undefined \def \url#1{\textsf{#1}}\fi
\ifx \bchapter \undefined \def \bchapter#1{#1}\fi
\ifx \bbook \undefined \def \bbook#1{#1}\fi
\ifx \bcomment \undefined \def \bcomment#1{#1}\fi
\ifx \oauthor \undefined \def \oauthor#1{#1}\fi
\ifx \citeauthoryear \undefined \def \citeauthoryear#1{#1}\fi
\ifx \endbibitem  \undefined \def \endbibitem {}\fi
\ifx \bconflocation  \undefined \def \bconflocation#1{#1}\fi
\ifx \arxivurl  \undefined \def \arxivurl#1{\textsf{#1}}\fi
\csname PreBibitemsHook\endcsname

\bibitem{ho_aamc}
\begin{botherref}
\oauthor{\bsnm{AAMC}}:
We can’t defeat COVID-19 without vaccinating children. There aren't even any
  kids' clinical trials yet
(2020).
\url{https://www.aamc.org/news-insights/we-can-t-defeat-covid-19-without-vaccinating-children-there-arent-even-any-kids-clinical-trials-yet}
\end{botherref}
\endbibitem

\bibitem{crooke2019immunosenescence}
\begin{barticle}
\bauthor{\bsnm{Crooke}, \binits{S.N.}},
\bauthor{\bsnm{Ovsyannikova}, \binits{I.G.}},
\bauthor{\bsnm{Poland}, \binits{G.A.}},
\bauthor{\bsnm{Kennedy}, \binits{R.B.}}:
\batitle{Immunosenescence and human vaccine immune responses}.
\bjtitle{Immunity \& Ageing}
\bvolume{16}(\bissue{1}),
\bfpage{25}
(\byear{2019})
\end{barticle}
\endbibitem

\bibitem{WHOvacc}
\begin{botherref}
\oauthor{\bsnm{WHO}}:
Draft landscape of covid 19 candidate vaccines
(2020).
\url{https://www.who.int/publications/m/item/draft-landscape-of-covid-19-candidate-vaccines}
\end{botherref}
\endbibitem

\bibitem{hu2021revealing}
\begin{botherref}
\oauthor{\bsnm{Hu}, \binits{T.}},
\oauthor{\bsnm{Wang}, \binits{S.}},
\oauthor{\bsnm{Luo}, \binits{W.}},
\oauthor{\bsnm{Yan}, \binits{Y.}},
\oauthor{\bsnm{Zhang}, \binits{M.}},
\oauthor{\bsnm{Huang}, \binits{X.}},
\oauthor{\bsnm{Liu}, \binits{R.}},
\oauthor{\bsnm{Ly}, \binits{K.}},
\oauthor{\bsnm{Kacker}, \binits{V.}},
\oauthor{\bsnm{Li}, \binits{Z.}}:
Revealing public opinion towards covid-19 vaccines using twitter data in the
  united states: a spatiotemporal perspective.
medRxiv
(2021)
\end{botherref}
\endbibitem

\bibitem{forman2021covid}
\begin{botherref}
\oauthor{\bsnm{Forman}, \binits{R.}},
\oauthor{\bsnm{Shah}, \binits{S.}},
\oauthor{\bsnm{Jeurissen}, \binits{P.}},
\oauthor{\bsnm{Jit}, \binits{M.}},
\oauthor{\bsnm{Mossialos}, \binits{E.}}:
Covid-19 vaccine challenges: What have we learned so far and what remains to be
  done?
Health Policy
(2021)
\end{botherref}
\endbibitem

\bibitem{jean2021vaccine}
\begin{barticle}
\bauthor{\bsnm{Jean-Jacques}, \binits{M.}},
\bauthor{\bsnm{Bauchner}, \binits{H.}}:
\batitle{Vaccine distribution—equity left behind?}
\bjtitle{JAMA}
\bvolume{325}(\bissue{9}),
\bfpage{829}--\blpage{830}
(\byear{2021})
\end{barticle}
\endbibitem

\bibitem{aborode2021equal}
\begin{botherref}
\oauthor{\bsnm{Aborode}, \binits{A.T.}},
\oauthor{\bsnm{Olofinsao}, \binits{O.A.}},
\oauthor{\bsnm{Osmond}, \binits{E.}},
\oauthor{\bsnm{Batubo}, \binits{A.P.}},
\oauthor{\bsnm{Fayemiro}, \binits{O.}},
\oauthor{\bsnm{Sherifdeen}, \binits{O.}},
\oauthor{\bsnm{Muraina}, \binits{L.}},
\oauthor{\bsnm{Obadawo}, \binits{B.}},
\oauthor{\bsnm{Ahmad}, \binits{S.}},
\oauthor{\bsnm{Fajemisin}, \binits{E.A.}}:
Equal access of covid-19 vaccine distribution in africa: Challenges and way
  forward.
Journal of Medical Virology
(2021)
\end{botherref}
\endbibitem

\bibitem{sharma2020review}
\begin{barticle}
\bauthor{\bsnm{Sharma}, \binits{O.}},
\bauthor{\bsnm{Sultan}, \binits{A.A.}},
\bauthor{\bsnm{Ding}, \binits{H.}},
\bauthor{\bsnm{Triggle}, \binits{C.R.}}:
\batitle{A review of the progress and challenges of developing a vaccine for
  covid-19}.
\bjtitle{Frontiers in immunology}
\bvolume{11},
\bfpage{2413}
(\byear{2020})
\end{barticle}
\endbibitem

\bibitem{newton2020covid}
\begin{barticle}
\bauthor{\bsnm{Newton}, \binits{P.N.}},
\bauthor{\bsnm{Bond}, \binits{K.C.}},
\bauthor{\bsnm{Adeyeye}, \binits{M.}},
\bauthor{\bsnm{Antignac}, \binits{M.}},
\bauthor{\bsnm{Ashenef}, \binits{A.}},
\bauthor{\bsnm{Awab}, \binits{G.R.}},
\bauthor{\bsnm{Bannenberg}, \binits{W.J.}},
\bauthor{\bsnm{Bower}, \binits{J.}},
\bauthor{\bsnm{Breman}, \binits{J.}},
\bauthor{\bsnm{Brock}, \binits{A.}}, \betal:
\batitle{Covid-19 and risks to the supply and quality of tests, drugs, and
  vaccines}.
\bjtitle{The Lancet Global Health}
\bvolume{8}(\bissue{6}),
\bfpage{754}--\blpage{755}
(\byear{2020})
\end{barticle}
\endbibitem

\bibitem{alam2021challenges}
\begin{barticle}
\bauthor{\bsnm{Alam}, \binits{S.T.}},
\bauthor{\bsnm{Ahmed}, \binits{S.}},
\bauthor{\bsnm{Ali}, \binits{S.M.}},
\bauthor{\bsnm{Sarker}, \binits{S.}},
\bauthor{\bsnm{Kabir}, \binits{G.}}, \betal:
\batitle{Challenges to covid-19 vaccine supply chain: Implications for
  sustainable development goals}.
\bjtitle{International Journal of Production Economics}
\bvolume{239},
\bfpage{108193}
(\byear{2021})
\end{barticle}
\endbibitem

\bibitem{NAP25917}
\begin{bchapter}
\bauthor{\bsnm{Kahn}, \binits{B.}},
\bauthor{\bsnm{Brown}, \binits{L.}},
\bauthor{\bsnm{Foege}, \binits{W.}},
\bauthor{\bsnm{Gayle}, \binits{H.}},
\bauthor{\bparticle{of} \bsnm{Sciences~Engineering}, \binits{N.A.}},
\bauthor{\bsnm{Medicine}}, \betal:
\bctitle{A framework for equitable allocation of covid-19 vaccine}.
In: \bbtitle{Framework for Equitable Allocation of COVID-19 Vaccine}.
\bpublisher{National Academies Press (US)}, \blocation{???}
(\byear{2020})
\end{bchapter}
\endbibitem

\bibitem{WHOfair}
\begin{botherref}
\oauthor{\bsnm{WHO}}:
Fair allocation mechanism for COVID-19 vaccines through the COVAX Facility
(2020).
\url{https://www.who.int/publications/m/item/fair-allocation-mechanism-for-covid-19-vaccines-through-the-covax-facility}
\end{botherref}
\endbibitem

\bibitem{CDCfair}
\begin{botherref}
\oauthor{\bsnm{CDC}}:
COVID-19 Vaccination Program Interim Playbook for jurisdiction Operations
(2020).
\url{https://www.cdc.gov/vaccines/imz-managers/downloads/COVID-19-Vaccination-Program-Interim_Playbook.pdf}
\end{botherref}
\endbibitem

\bibitem{clinicaltrials.gov}
\begin{botherref}
Novelty, conformity and trust in vaccines - full text view.
\url{https://clinicaltrials.gov/ct2/show/NCT04693689}
\end{botherref}
\endbibitem

\bibitem{HHS}
\begin{botherref}
\oauthor{\bsnm{HHS}}:
Explaining operation warp speed
(2020).
\url{https://www.hhs.gov/coronavirus/explaining-operation-warp-speed/index.html}
\end{botherref}
\endbibitem

\bibitem{GAVI}
\begin{botherref}
\oauthor{\bsnm{GAVI}}:
Covax Explained
(2020).
\url{https://www.gavi.org/vaccineswork/covax-explained}
\end{botherref}
\endbibitem

\bibitem{modernapress}
\begin{botherref}
\oauthor{\bsnm{Moderna}}:
Moderna\'s covid-19 vaccine candidate meets its primary efficacy
(2020).
\url{https://investors.modernatx.com/news-releases/news-release-details/modernas-covid-19-vaccine-candidate-meets-its-primary-efficacy}
\end{botherref}
\endbibitem

\bibitem{pfizerpress}
\begin{botherref}
\oauthor{\bsnm{pfizer}}:
pfizer and biontech conclude phase 3 study covid-19 vaccine
(2020).
\url{https://www.pfizer.com/news/press-release/press-release-detail/pfizer-and-biontech-conclude-phase-3-study-covid-19-vaccine}
\end{botherref}
\endbibitem

\bibitem{joebidencombatplan}
\begin{botherref}
\oauthor{\bsnm{Biden}, \binits{J.}}:
THE BIDEN PLAN TO COMBAT CORONAVIRUS (COVID-19) AND PREPARE FOR FUTURE GLOBAL
  HEALTH THREATS
(2020).
\url{https://joebiden.com/covid-plan/}
\end{botherref}
\endbibitem

\bibitem{joebidencovid19}
\begin{botherref}
\oauthor{\bsnm{Biden}, \binits{J.}}:
JOE AND KAMALA'S PLAN TO BEAT COVID-19
(2020).
\url{https://joebiden.com/covid19/}
\end{botherref}
\endbibitem

\bibitem{CDC}
\begin{botherref}
\oauthor{\bsnm{CDC}}:
How the Flu Virus Can Change: “Drift” and “Shift”
(2020).
\url{https://www.cdc.gov/flu/about/viruses/change.htm}
\end{botherref}
\endbibitem

\bibitem{boni_2008}
\begin{botherref}
\oauthor{\bsnm{Boni}, \binits{M.F.}}:
Vaccination and antigenic drift in influenza.
Vaccine
\textbf{26}
(2008).
doi:\doiurl{10.1016/j.vaccine.2008.04.011}
\end{botherref}
\endbibitem

\bibitem{cdch1n1}
\begin{botherref}
Guidance from Pediatric Stakeholders: A Coordinated Approach to Communicating
  Pediatric-related Information on Pandemic Influenza at the Community Level.
\url{https://www.cdc.gov/h1n1flu/guidance/pediatrics_tool.htm}
\end{botherref}
\endbibitem

\bibitem{fineberg2014pandemic}
\begin{barticle}
\bauthor{\bsnm{Fineberg}, \binits{H.V.}}:
\batitle{Pandemic preparedness and response—lessons from the h1n1 influenza
  of 2009}.
\bjtitle{New England Journal of Medicine}
\bvolume{370}(\bissue{14}),
\bfpage{1335}--\blpage{1342}
(\byear{2014})
\end{barticle}
\endbibitem

\bibitem{vousden2020lessons}
\begin{botherref}
\oauthor{\bsnm{Vousden}, \binits{N.}},
\oauthor{\bsnm{Knight}, \binits{M.}}:
Lessons learned from the a (h1n1) influenza pandemic.
Best Practice \& Research Clinical Obstetrics \& Gynaecology
(2020)
\end{botherref}
\endbibitem

\bibitem{GAO}
\begin{botherref}
\oauthor{\bsnm{{U.S. Government Accountability Office}}}:
COVID-19: Federal Efforts Accelerate Vaccine and Therapeutic Development, but
  More Transparency Needed on Emergency Use Authorizations
(2020).
\url{https://www.gao.gov/products/GAO-21-207}
\end{botherref}
\endbibitem

\bibitem{cdc_acip_2020}
\begin{botherref}
\oauthor{\bsnm{{CDC}}}:
{ACIP} December 2020 Presentation Slides {\textbar} Immunization Practices
  {\textbar} {CDC}.
\url{https://www.cdc.gov/vaccines/acip/meetings/slides-2020-12.html}
Accessed 2020-12-03
\end{botherref}
\endbibitem

\bibitem{distribution_nys}
\begin{botherref}
\oauthor{\bsnm{State}, \binits{N.Y.}}:
National Governors Association Submits List of Questions to Trump
  Administration on Effective Implementation of COVID-19 Vaccine
(2020).
\url{https://www.governor.ny.gov/news/national-governors-association-submits-list-questions-trump-administration-effective}
\end{botherref}
\endbibitem

\bibitem{distribution_nyt}
\begin{botherref}
\oauthor{\bsnm{Times}, \binits{N.Y.}}:
How to Ship a Vaccine at $\sim$80°C, and Other Obstacles in the Covid Fight
(2020).
\url{https://www.nytimes.com/2020/09/18/business/coronavirus-covid-vaccine-cold-frozen-logistics.html}
\end{botherref}
\endbibitem

\bibitem{distribution_nh}
\begin{botherref}
\oauthor{\bsnm{Herald}, \binits{N.}}:
COVID-19: The logistic challenges of vaccine distribution in India
(2020).
\url{https://www.nationalheraldindia.com/india/covid-19-the-logistic-challenges-of-vaccine-distribution-in-india}
\end{botherref}
\endbibitem

\bibitem{distribution_wp}
\begin{botherref}
\oauthor{\bsnm{Washingtonpost}}:
The biggest problem isn’t when a vaccine is ready. It’s how to distribute
  it
(2020).
\url{https://www.washingtonpost.com/outlook/2020/10/27/coronavirus-vaccine-distribution-plan/}
\end{botherref}
\endbibitem

\bibitem{mugali2017improving}
\begin{barticle}
\bauthor{\bsnm{Mugali}, \binits{R.R.}},
\bauthor{\bsnm{Mansoor}, \binits{F.}},
\bauthor{\bsnm{Parwiz}, \binits{S.}},
\bauthor{\bsnm{Ahmad}, \binits{F.}},
\bauthor{\bsnm{Safi}, \binits{N.}},
\bauthor{\bsnm{Higgins-Steele}, \binits{A.}},
\bauthor{\bsnm{Varkey}, \binits{S.}}:
\batitle{Improving immunization in afghanistan: results from a cross-sectional
  community-based survey to assess routine immunization coverage}.
\bjtitle{BMC Public Health}
\bvolume{17}(\bissue{1}),
\bfpage{290}
(\byear{2017})
\end{barticle}
\endbibitem

\bibitem{distribution_worldbank}
\begin{botherref}
\oauthor{\bsnm{Bank}, \binits{T.W.}}:
Country Engagement
(2020).
\url{https://id4d.worldbank.org/country-engagement}
\end{botherref}
\endbibitem

\bibitem{distribution_la_times}
\begin{botherref}
\oauthor{\bsnm{Times}, \binits{L.A.}}:
H1N1 vaccine was unevenly distributed across L.A. County, figures show
(2010).
\url{https://www.latimes.com/archives/la-xpm-2010-mar-01-la-me-h1n1-supply1-2010mar01-story.html}
\end{botherref}
\endbibitem

\bibitem{masia2018vaccination}
\begin{barticle}
\bauthor{\bsnm{Masia}, \binits{N.A.}},
\bauthor{\bsnm{Smerling}, \binits{J.}},
\bauthor{\bsnm{Kapfidze}, \binits{T.}},
\bauthor{\bsnm{Manning}, \binits{R.}},
\bauthor{\bsnm{Showalter}, \binits{M.}}:
\batitle{Vaccination and gdp growth rates: Exploring the links in a conditional
  convergence framework}.
\bjtitle{World Development}
\bvolume{103},
\bfpage{88}--\blpage{99}
(\byear{2018})
\end{barticle}
\endbibitem

\bibitem{distribution_cdc}
\begin{botherref}
\oauthor{\bsnm{CDC}}:
U.S. Public Health Response to the Measles Outbreak
(2019).
\url{https://www.cdc.gov/washington/testimony/2019/t20190227.htm}
\end{botherref}
\endbibitem

\bibitem{fauci2014immune}
\begin{barticle}
\bauthor{\bsnm{Fauci}, \binits{A.S.}},
\bauthor{\bsnm{Marovich}, \binits{M.A.}},
\bauthor{\bsnm{Dieffenbach}, \binits{C.W.}},
\bauthor{\bsnm{Hunter}, \binits{E.}},
\bauthor{\bsnm{Buchbinder}, \binits{S.P.}}:
\batitle{Immune activation with hiv vaccines}.
\bjtitle{Science}
\bvolume{344}(\bissue{6179}),
\bfpage{49}--\blpage{51}
(\byear{2014})
\end{barticle}
\endbibitem

\bibitem{white_could_2020}
\begin{botherref}
\oauthor{\bsnm{White}, \binits{S.}}:
Could {COVID}-19 {mRNA} vaccines cause autoimmune diseases?
BMJ
(2020).
Accessed 2020-12-03
\end{botherref}
\endbibitem

\bibitem{buchbinder2020use}
\begin{botherref}
\oauthor{\bsnm{Buchbinder}, \binits{S.P.}},
\oauthor{\bsnm{McElrath}, \binits{M.J.}},
\oauthor{\bsnm{Dieffenbach}, \binits{C.}},
\oauthor{\bsnm{Corey}, \binits{L.}}:
Use of adenovirus type-5 vectored vaccines: a cautionary tale.
The Lancet
(2020)
\end{botherref}
\endbibitem

\bibitem{rauch2018new}
\begin{barticle}
\bauthor{\bsnm{Rauch}, \binits{S.}},
\bauthor{\bsnm{Jasny}, \binits{E.}},
\bauthor{\bsnm{Schmidt}, \binits{K.E.}},
\bauthor{\bsnm{Petsch}, \binits{B.}}:
\batitle{New vaccine technologies to combat outbreak situations}.
\bjtitle{Frontiers in immunology}
\bvolume{9},
\bfpage{1963}
(\byear{2018})
\end{barticle}
\endbibitem

\bibitem{pardi2018mrna}
\begin{barticle}
\bauthor{\bsnm{Pardi}, \binits{N.}},
\bauthor{\bsnm{Hogan}, \binits{M.J.}},
\bauthor{\bsnm{Porter}, \binits{F.W.}},
\bauthor{\bsnm{Weissman}, \binits{D.}}:
\batitle{mrna vaccines—a new era in vaccinology}.
\bjtitle{Nature reviews Drug discovery}
\bvolume{17}(\bissue{4}),
\bfpage{261}
(\byear{2018})
\end{barticle}
\endbibitem

\bibitem{ho_j_and_j}
\begin{botherref}
\oauthor{\bsnm{Johnson}, \binits{J..}}:
Johnson \& Johnson Prepares to Resume Phase 3 ENSEMBLE Trial of its Janssen
  COVID-19 Vaccine Candidate in the U.S.
(2020).
\url{https://www.jnj.com/our-company/johnson-johnson-prepares-to-resume-phase-3-ensemble-trial-of-its-janssen-covid-19-vaccine-candidate-in-the-us}
\end{botherref}
\endbibitem

\bibitem{ho_astrazeneca}
\begin{botherref}
\oauthor{\bsnm{Astrazeneca}}:
COVID-19 vaccine AZD1222 clinical trials resumed in the UK
(2020).
\url{https://www.astrazeneca.com/media-centre/press-releases/2020/covid-19-vaccine-azd1222-clinical-trials-resumed-in-the-uk.html}
\end{botherref}
\endbibitem

\bibitem{thacker2009strategies}
\begin{barticle}
\bauthor{\bsnm{Thacker}, \binits{E.E.}},
\bauthor{\bsnm{Timares}, \binits{L.}},
\bauthor{\bsnm{Matthews}, \binits{Q.L.}}:
\batitle{Strategies to overcome host immunity to adenovirus vectors in vaccine
  development}.
\bjtitle{Expert review of vaccines}
\bvolume{8}(\bissue{6}),
\bfpage{761}--\blpage{777}
(\byear{2009})
\end{barticle}
\endbibitem

\bibitem{ho_cen}
\begin{botherref}
\oauthor{\bsnm{C\&EN}}:
Adenoviral vectors are the new COVID-19 vaccine front-runners. Can they
  overcome their checkered past?
(2020).
\url{https://cen.acs.org/pharmaceuticals/vaccines/Adenoviral-vectors-new-COVID-19/98/i19}
\end{botherref}
\endbibitem

\bibitem{zheng2020multiple}
\begin{barticle}
\bauthor{\bsnm{Zheng}, \binits{C.}},
\bauthor{\bsnm{Kar}, \binits{I.}},
\bauthor{\bsnm{Chen}, \binits{C.K.}},
\bauthor{\bsnm{Sau}, \binits{C.}},
\bauthor{\bsnm{Woodson}, \binits{S.}},
\bauthor{\bsnm{Serra}, \binits{A.}},
\bauthor{\bsnm{Abboud}, \binits{H.}}:
\batitle{Multiple sclerosis disease-modifying therapy and the covid-19
  pandemic: implications on the risk of infection and future vaccination}.
\bjtitle{CNS drugs}
\bvolume{34}(\bissue{9}),
\bfpage{879}--\blpage{896}
(\byear{2020})
\end{barticle}
\endbibitem

\bibitem{edridge2020seasonal}
\begin{botherref}
\oauthor{\bsnm{Edridge}, \binits{A.W.}},
\oauthor{\bsnm{Kaczorowska}, \binits{J.}},
\oauthor{\bsnm{Hoste}, \binits{A.C.}},
\oauthor{\bsnm{Bakker}, \binits{M.}},
\oauthor{\bsnm{Klein}, \binits{M.}},
\oauthor{\bsnm{Loens}, \binits{K.}},
\oauthor{\bsnm{Jebbink}, \binits{M.F.}},
\oauthor{\bsnm{Matser}, \binits{A.}},
\oauthor{\bsnm{Kinsella}, \binits{C.M.}},
\oauthor{\bsnm{Rueda}, \binits{P.}}, et al.:
Seasonal coronavirus protective immunity is short-lasting.
Nature medicine,
1--3
(2020)
\end{botherref}
\endbibitem

\bibitem{dan2020immunological}
\begin{botherref}
\oauthor{\bsnm{Dan}, \binits{J.M.}},
\oauthor{\bsnm{Mateus}, \binits{J.}},
\oauthor{\bsnm{Kato}, \binits{Y.}},
\oauthor{\bsnm{Hastie}, \binits{K.M.}},
\oauthor{\bsnm{Faliti}, \binits{C.}},
\oauthor{\bsnm{Ramirez}, \binits{S.I.}},
\oauthor{\bsnm{Frazier}, \binits{A.}},
\oauthor{\bsnm{Esther}, \binits{D.Y.}},
\oauthor{\bsnm{Grifoni}, \binits{A.}},
\oauthor{\bsnm{Rawlings}, \binits{S.A.}}, et al.:
Immunological memory to sars-cov-2 assessed for greater than six months after
  infection.
bioRxiv
(2020)
\end{botherref}
\endbibitem

\bibitem{cruz2021duration}
\begin{botherref}
\oauthor{\bsnm{Cruz}, \binits{A.T.}},
\oauthor{\bsnm{Zeichner}, \binits{S.L.}}:
Duration of effective antibody levels after covid-19.
Pediatrics
\textbf{148}(3)
(2021)
\end{botherref}
\endbibitem

\bibitem{ortega2021seven}
\begin{barticle}
\bauthor{\bsnm{Ortega}, \binits{N.}},
\bauthor{\bsnm{Ribes}, \binits{M.}},
\bauthor{\bsnm{Vidal}, \binits{M.}},
\bauthor{\bsnm{Rubio}, \binits{R.}},
\bauthor{\bsnm{Aguilar}, \binits{R.}},
\bauthor{\bsnm{Williams}, \binits{S.}},
\bauthor{\bsnm{Barrios}, \binits{D.}},
\bauthor{\bsnm{Alonso}, \binits{S.}},
\bauthor{\bsnm{Hern{\'a}ndez-Luis}, \binits{P.}},
\bauthor{\bsnm{Mitchell}, \binits{R.A.}}, \betal:
\batitle{Seven-month kinetics of sars-cov-2 antibodies and role of pre-existing
  antibodies to human coronaviruses}.
\bjtitle{Nature Communications}
\bvolume{12}(\bissue{1}),
\bfpage{1}--\blpage{10}
(\byear{2021})
\end{barticle}
\endbibitem

\bibitem{wendelboe2005duration}
\begin{barticle}
\bauthor{\bsnm{Wendelboe}, \binits{A.M.}},
\bauthor{\bsnm{Van~Rie}, \binits{A.}},
\bauthor{\bsnm{Salmaso}, \binits{S.}},
\bauthor{\bsnm{Englund}, \binits{J.A.}}:
\batitle{Duration of immunity against pertussis after natural infection or
  vaccination}.
\bjtitle{The Pediatric infectious disease journal}
\bvolume{24}(\bissue{5}),
\bfpage{58}--\blpage{61}
(\byear{2005})
\end{barticle}
\endbibitem

\bibitem{sah2018optimizing}
\begin{barticle}
\bauthor{\bsnm{Sah}, \binits{P.}},
\bauthor{\bsnm{Medlock}, \binits{J.}},
\bauthor{\bsnm{Fitzpatrick}, \binits{M.C.}},
\bauthor{\bsnm{Singer}, \binits{B.H.}},
\bauthor{\bsnm{Galvani}, \binits{A.P.}}:
\batitle{Optimizing the impact of low-efficacy influenza vaccines}.
\bjtitle{Proceedings of the National Academy of Sciences}
\bvolume{115}(\bissue{20}),
\bfpage{5151}--\blpage{5156}
(\byear{2018})
\end{barticle}
\endbibitem

\bibitem{okoli2021decline}
\begin{bchapter}
\bauthor{\bsnm{Okoli}, \binits{G.N.}},
\bauthor{\bsnm{Racovitan}, \binits{F.}},
\bauthor{\bsnm{Abdulwahid}, \binits{T.}},
\bauthor{\bsnm{Hyder}, \binits{S.K.}},
\bauthor{\bsnm{Lansbury}, \binits{L.}},
\bauthor{\bsnm{Righolt}, \binits{C.H.}},
\bauthor{\bsnm{Mahmud}, \binits{S.M.}},
\bauthor{\bsnm{Nguyen-Van-Tam}, \binits{J.S.}}:
\bctitle{Decline in seasonal influenza vaccine effectiveness with vaccination
  program maturation: A systematic review and meta-analysis}.
In: \bbtitle{Open Forum Infectious Diseases},
vol. \bseriesno{8},
p. \bfpage{069}
(\byear{2021}).
\bcomment{Oxford University Press US}
\end{bchapter}
\endbibitem

\bibitem{ho_pfizer}
\begin{botherref}
\oauthor{\bsnm{Pfizer}}:
OUR PROGRESS IN DEVELOPING A POTENTIAL COVID-19 VACCINE
(2020).
\url{https://www.pfizer.com/science/coronavirus/vaccine}
\end{botherref}
\endbibitem

\bibitem{helfand2020exclusion}
\begin{botherref}
\oauthor{\bsnm{Helfand}, \binits{B.K.}},
\oauthor{\bsnm{Webb}, \binits{M.}},
\oauthor{\bsnm{Gartaganis}, \binits{S.L.}},
\oauthor{\bsnm{Fuller}, \binits{L.}},
\oauthor{\bsnm{Kwon}, \binits{C.-S.}},
\oauthor{\bsnm{Inouye}, \binits{S.K.}}:
The exclusion of older persons from vaccine and treatment trials for
  coronavirus disease 2019—missing the target.
JAMA Internal Medicine
(2020)
\end{botherref}
\endbibitem

\bibitem{ho_fda}
\begin{botherref}
\oauthor{\bsnm{FDA}}:
Development and Licensure of Vaccines to Prevent COVID-19 Guidance for Industry
(2020).
\url{https://www.fda.gov/media/139638/download}
\end{botherref}
\endbibitem

\bibitem{YANG20203184}
\begin{barticle}
\bauthor{\bsnm{Yang}, \binits{Y.T.}},
\bauthor{\bsnm{Pendo}, \binits{E.}},
\bauthor{\bsnm{Reiss}, \binits{D.R.}}:
\batitle{The americans with disabilities act and healthcare employer-mandated
  vaccinations}.
\bjtitle{Vaccine}
\bvolume{38}(\bissue{16}),
\bfpage{3184}--\blpage{3186}
(\byear{2020}).
doi:\doiurl{10.1016/j.vaccine.2020.03.012"}
\end{barticle}
\endbibitem

\bibitem{ho_pew}
\begin{botherref}
\oauthor{\bsnm{Research}, \binits{P.}}:
U.S. Public Now Divided Over Whether To Get COVID-19 Vaccine
(2020).
\url{https://www.pewresearch.org/science/2020/09/17/u-s-public-now-divided-over-whether-to-get-covid-19-vaccine/}
\end{botherref}
\endbibitem

\bibitem{ho_cnn}
\begin{botherref}
\oauthor{\bsnm{CNN}}:
STATE OF THE UNION
(2020).
\url{http://transcripts.cnn.com/TRANSCRIPTS/2011/15/sotu.01.html}
\end{botherref}
\endbibitem

\bibitem{ho_cdc_freq}
\begin{botherref}
\oauthor{\bsnm{CDC}}:
Frequently Asked Questions about COVID-19 Vaccination
(2020).
\url{https://www.cdc.gov/coronavirus/2019-ncov/vaccines/faq.html}
\end{botherref}
\endbibitem

\bibitem{lazarus2020global}
\begin{botherref}
\oauthor{\bsnm{Lazarus}, \binits{J.V.}},
\oauthor{\bsnm{Ratzan}, \binits{S.C.}},
\oauthor{\bsnm{Palayew}, \binits{A.}},
\oauthor{\bsnm{Gostin}, \binits{L.O.}},
\oauthor{\bsnm{Larson}, \binits{H.J.}},
\oauthor{\bsnm{Rabin}, \binits{K.}},
\oauthor{\bsnm{Kimball}, \binits{S.}},
\oauthor{\bsnm{El-Mohandes}, \binits{A.}}:
A global survey of potential acceptance of a covid-19 vaccine.
Nature medicine,
1--4
(2020)
\end{botherref}
\endbibitem

\bibitem{uci_who}
\begin{botherref}
\oauthor{\bsnm{WHO}}:
Draft landscape of COVID-19 candidate vaccines
(2020).
\url{https://www.who.int/publications/m/item/draft-landscape-of-covid-19-candidate-vaccines}
\end{botherref}
\endbibitem

\bibitem{walsh2020safety}
\begin{botherref}
\oauthor{\bsnm{Walsh}, \binits{E.E.}},
\oauthor{\bsnm{Frenck~Jr}, \binits{R.W.}},
\oauthor{\bsnm{Falsey}, \binits{A.R.}},
\oauthor{\bsnm{Kitchin}, \binits{N.}},
\oauthor{\bsnm{Absalon}, \binits{J.}},
\oauthor{\bsnm{Gurtman}, \binits{A.}},
\oauthor{\bsnm{Lockhart}, \binits{S.}},
\oauthor{\bsnm{Neuzil}, \binits{K.}},
\oauthor{\bsnm{Mulligan}, \binits{M.J.}},
\oauthor{\bsnm{Bailey}, \binits{R.}}, et al.:
Safety and immunogenicity of two rna-based covid-19 vaccine candidates.
New England Journal of Medicine
(2020)
\end{botherref}
\endbibitem

\bibitem{graham2020evaluation}
\begin{barticle}
\bauthor{\bsnm{Graham}, \binits{S.P.}},
\bauthor{\bsnm{McLean}, \binits{R.K.}},
\bauthor{\bsnm{Spencer}, \binits{A.J.}},
\bauthor{\bsnm{Belij-Rammerstorfer}, \binits{S.}},
\bauthor{\bsnm{Wright}, \binits{D.}},
\bauthor{\bsnm{Ulaszewska}, \binits{M.}},
\bauthor{\bsnm{Edwards}, \binits{J.C.}},
\bauthor{\bsnm{Hayes}, \binits{J.W.}},
\bauthor{\bsnm{Martini}, \binits{V.}},
\bauthor{\bsnm{Thakur}, \binits{N.}}, \betal:
\batitle{Evaluation of the immunogenicity of prime-boost vaccination with the
  replication-deficient viral vectored covid-19 vaccine candidate chadox1
  ncov-19}.
\bjtitle{NPJ vaccines}
\bvolume{5}(\bissue{1}),
\bfpage{1}--\blpage{6}
(\byear{2020})
\end{barticle}
\endbibitem

\bibitem{uci_cdc2}
\begin{botherref}
\oauthor{\bsnm{CDC}}:
PanVax Tool for Pandemic Vaccination Planning
(2020).
\url{https://www.cdc.gov/flu/pandemic-resources/tools/panvax-tool.htm}
\end{botherref}
\endbibitem

\bibitem{uci_cdc3}
\begin{botherref}
\oauthor{\bsnm{CDC}}:
Vaccine Tracking System (VTrckS)
(2020).
\url{https://www.cdc.gov/vaccines/programs/vtrcks/index.html}
\end{botherref}
\endbibitem

\bibitem{uci_palantir}
\begin{botherref}
\oauthor{\bsnm{Journal}, \binits{T.W.S.}}:
Palantir to Help U.S. Track Covid-19 Vaccines
(2020).
\url{https://www.wsj.com/articles/palantir-to-help-u-s-track-covid-19-vaccines-11603367276}
\end{botherref}
\endbibitem

\bibitem{uci_sap}
\begin{botherref}
\oauthor{\bsnm{SAP}}:
Vaccine Collaboration Hub from SAP Improves Supply Chain Efficiency for
  Government and Life Sciences Organizations
(2020).
\url{https://news.sap.com/2020/10/vaccine-collaboration-hub-supply-chain-efficiency-government-life-sciences/}
\end{botherref}
\endbibitem

\bibitem{uci_accenture}
\begin{botherref}
\oauthor{\bsnm{Accenture}}:
Accenture Vaccine Management Solution
(2020).
\url{https://www.accenture.com/us-en/services/public-service/vaccine-management-solution}
\end{botherref}
\endbibitem

\bibitem{uci_maryland}
\begin{botherref}
\oauthor{\bparticle{of} \bsnm{Health}, \binits{M.D.}}:
COVID-19 Vaccination Plan
(2020).
\url{https://phpa.health.maryland.gov/Documents/10.19.2020_Maryland_COVID-19_Vaccination_Plan_CDCwm.pdf}
\end{botherref}
\endbibitem

\bibitem{uci_bookings}
\begin{botherref}
\oauthor{\bsnm{Bookings}}:
Want herd immunity? Pay people to take the vaccine
(2020).
\url{https://www.brookings.edu/opinions/want-herd-immunity-pay-people-to-take-the-vaccine/}
\end{botherref}
\endbibitem

\bibitem{enact}
\begin{bchapter}
\bauthor{\bsnm{Prasad}, \binits{A.}},
\bauthor{\bsnm{Kotz}, \binits{D.}}:
\bctitle{Enact: Encounter-based architecture for contact tracing}.
In: \bbtitle{Proceedings of the 4th International on Workshop on Physical
  Analytics}.
\bsertitle{WPA '17},
pp. \bfpage{37}--\blpage{42}.
\bpublisher{Association for Computing Machinery},
\blocation{New York, NY, USA}
(\byear{2017}).
doi:\doiurl{10.1145/3092305.3092310}.
\burl{https://doi.org/10.1145/3092305.3092310}
\end{bchapter}
\endbibitem

\bibitem{epione}
\begin{botherref}
\oauthor{\bsnm{Trieu}, \binits{N.}},
\oauthor{\bsnm{Shehata}, \binits{K.}},
\oauthor{\bsnm{Saxena}, \binits{P.}},
\oauthor{\bsnm{Shokri}, \binits{R.}},
\oauthor{\bsnm{Song}, \binits{D.}}:
Epione: Lightweight Contact Tracing with Strong Privacy
(2020).
\arxivurl{2004.13293}
\end{botherref}
\endbibitem

\bibitem{Rogue}
\begin{botherref}
\oauthor{\bsnm{Raskar}, \binits{R.}},
\oauthor{\bsnm{Schunemann}, \binits{I.}},
\oauthor{\bsnm{Barbar}, \binits{R.}},
\oauthor{\bsnm{Vilcans}, \binits{K.}},
\oauthor{\bsnm{Gray}, \binits{J.}},
\oauthor{\bsnm{Vepakomma}, \binits{P.}},
\oauthor{\bsnm{Kapa}, \binits{S.}},
\oauthor{\bsnm{Nuzzo}, \binits{A.}},
\oauthor{\bsnm{Gupta}, \binits{R.}},
\oauthor{\bsnm{Berke}, \binits{A.}},
\oauthor{\bsnm{Greenwood}, \binits{D.}},
\oauthor{\bsnm{Keegan}, \binits{C.}},
\oauthor{\bsnm{Kanaparti}, \binits{S.}},
\oauthor{\bsnm{Beaudry}, \binits{R.}},
\oauthor{\bsnm{Stansbury}, \binits{D.}},
\oauthor{\bsnm{Arcila}, \binits{B.B.}},
\oauthor{\bsnm{Kanaparti}, \binits{R.}},
\oauthor{\bsnm{Pamplona}, \binits{V.}},
\oauthor{\bsnm{Benedetti}, \binits{F.M.}},
\oauthor{\bsnm{Clough}, \binits{A.}},
\oauthor{\bsnm{Das}, \binits{R.}},
\oauthor{\bsnm{Jain}, \binits{K.}},
\oauthor{\bsnm{Louisy}, \binits{K.}},
\oauthor{\bsnm{Nadeau}, \binits{G.}},
\oauthor{\bsnm{Pamplona}, \binits{V.}},
\oauthor{\bsnm{Penrod}, \binits{S.}},
\oauthor{\bsnm{Rajaee}, \binits{Y.}},
\oauthor{\bsnm{Singh}, \binits{A.}},
\oauthor{\bsnm{Storm}, \binits{G.}},
\oauthor{\bsnm{Werner}, \binits{J.}}:
Apps Gone Rogue: Maintaining Personal Privacy in an Epidemic
(2020).
\arxivurl{2003.08567}
\end{botherref}
\endbibitem

\bibitem{berke2020assessing}
\begin{botherref}
\oauthor{\bsnm{Berke}, \binits{A.}},
\oauthor{\bsnm{Bakker}, \binits{M.}},
\oauthor{\bsnm{Vepakomma}, \binits{P.}},
\oauthor{\bsnm{Larson}, \binits{K.}},
\oauthor{\bsnm{Pentland}, \binits{A.S.}}:
Assessing Disease Exposure Risk with Location Data: A Proposal for
  Cryptographic Preservation of Privacy
(2020).
\arxivurl{2003.14412}
\end{botherref}
\endbibitem

\bibitem{bell2020tracesecure}
\begin{botherref}
\oauthor{\bsnm{Bell}, \binits{J.}},
\oauthor{\bsnm{Butler}, \binits{D.}},
\oauthor{\bsnm{Hicks}, \binits{C.}},
\oauthor{\bsnm{Crowcroft}, \binits{J.}}:
TraceSecure: Towards Privacy Preserving Contact Tracing
(2020).
\arxivurl{2004.04059}
\end{botherref}
\endbibitem

\bibitem{comm_bjm_1}
\begin{botherref}
\oauthor{\bsnm{BJM}, \binits{T.}}:
Political interference in public health science during covid-19
(2020).
\url{https://www.bmj.com/content/371/bmj.m3878}
\end{botherref}
\endbibitem

\bibitem{comm_bjm_2}
\begin{botherref}
\oauthor{\bsnm{BJM}, \binits{T.}}:
Political interference in public health science during covid-19
(2020).
\url{https://www.bmj.com/content/371/bmj.m3878/rapid-responses}
\end{botherref}
\endbibitem

\bibitem{comm_et_healthworld}
\begin{botherref}
\oauthor{\bsnm{Healthworld}, \binits{E.}}:
Hydroxychloroquine promoted by Trump as Covid 'game changer' linked to
  increased deaths
(2020).
\url{https://health.economictimes.indiatimes.com/news/diagnostics/hydroxychloroquine-promoted-by-trump-as-covid-game-changer-linked-to-increased-deaths/75785673}
\end{botherref}
\endbibitem

\bibitem{comm_fauci}
\begin{botherref}
\oauthor{\bsnm{CNBC}}:
Dr. Fauci says all the ‘valid’ scientific data shows hydroxychloroquine
  isn’t effective in treating coronavirus
(2020).
\url{https://www.cnbc.com/2020/07/29/dr-fauci-says-all-the-valid-scientific-data-shows-hydroxychloroquine-isnt-effective-in-treating-coronavirus.html}
\end{botherref}
\endbibitem

\bibitem{comm_who}
\begin{botherref}
\oauthor{\bsnm{WHO}}:
More than 150 countries engaged in COVID-19 vaccine global access facility
(2020).
\url{https://www.who.int/news/item/15-07-2020-more-than-150-countries-engaged-in-covid-19-vaccine-global-access-facility}
\end{botherref}
\endbibitem

\bibitem{comm_stat}
\begin{botherref}
\oauthor{\bsnm{News}, \binits{S.}}:
STAT-Harris Poll: The share of Americans interested in getting Covid-19 vaccine
  as soon as possible is dropping
(2020).
\url{https://www.statnews.com/pharmalot/2020/10/19/covid19-coronavirus-pandemic-vaccine-racial-disparities/}
\end{botherref}
\endbibitem

\bibitem{comm_hastings}
\begin{botherref}
\oauthor{\bsnm{Center}, \binits{T.H.}}:
Public Trust in Science with Dr. Anthony Fauci
(2020).
\url{https://www.youtube.com/watch?v=Az1kD5xnzS4&ab_channel=TheHastingsCenter}
\end{botherref}
\endbibitem

\bibitem{comm_cidrap}
\begin{botherref}
\oauthor{\bsnm{CIDRAP}}:
Who will accept a COVID-19 vaccine?
(2020).
\url{https://www.cidrap.umn.edu/news-perspective/2020/10/who-will-accept-covid-19-vaccine}
\end{botherref}
\endbibitem

\bibitem{comm_stat2}
\begin{botherref}
\oauthor{\bsnm{News}, \binits{S.}}:
Health experts want to prioritize people of color for a Covid-19 vaccine. But
  how should it be done?
(2020).
\url{https://www.statnews.com/2020/11/09/health-experts-want-to-prioritize-people-of-color-for-covid19-vaccine-but-how-should-it-be-done/?utm_sourceSTAT+Newsletters&utm_campaign=00a79720d8-Daily_Recap&utm_medium=email&utm_term=0_8cab1d7961-00a79720d8-152583781}
\end{botherref}
\endbibitem

\bibitem{schmidt2020lawful}
\begin{botherref}
\oauthor{\bsnm{Schmidt}, \binits{H.}},
\oauthor{\bsnm{Gostin}, \binits{L.O.}},
\oauthor{\bsnm{Williams}, \binits{M.A.}}:
Is it lawful and ethical to prioritize racial minorities for covid-19 vaccines?
JAMA
(2020)
\end{botherref}
\endbibitem

\bibitem{patwa2020fighting}
\begin{botherref}
\oauthor{\bsnm{Patwa}, \binits{P.}},
\oauthor{\bsnm{Sharma}, \binits{S.}},
\oauthor{\bsnm{PYKL}, \binits{S.}},
\oauthor{\bsnm{Guptha}, \binits{V.}},
\oauthor{\bsnm{Kumari}, \binits{G.}},
\oauthor{\bsnm{Akhtar}, \binits{M.S.}},
\oauthor{\bsnm{Ekbal}, \binits{A.}},
\oauthor{\bsnm{Das}, \binits{A.}},
\oauthor{\bsnm{Chakraborty}, \binits{T.}}:
Fighting an Infodemic: COVID-19 Fake News Dataset
(2020).
\arxivurl{2011.03327}
\end{botherref}
\endbibitem

\bibitem{comm_timesofisrael}
\begin{botherref}
\oauthor{\bparticle{of} \bsnm{Israel}, \binits{T.}}:
Hundreds die of poisoning in Iran as fake news suggests methanol cure for virus
(2020).
\url{https://www.timesofisrael.com/hundreds-die-of-poisoning-in-iran-as-fake-news-suggests-methanol-cure-for-virus/}
\end{botherref}
\endbibitem

\bibitem{comm_bbc}
\begin{botherref}
\oauthor{\bsnm{BBC}}:
Coronavirus: Bill Gates ‘microchip’ conspiracy theory and other vaccine
  claims fact-checked
(2020).
\url{https://www.bbc.com/news/52847648}
\end{botherref}
\endbibitem

\bibitem{burki2019vaccine}
\begin{barticle}
\bauthor{\bsnm{Burki}, \binits{T.}}:
\batitle{Vaccine misinformation and social media}.
\bjtitle{The Lancet Digital Health}
\bvolume{1}(\bissue{6}),
\bfpage{258}--\blpage{259}
(\byear{2019})
\end{barticle}
\endbibitem

\bibitem{comm_bbc2}
\begin{botherref}
\oauthor{\bsnm{BBC}}:
Stella Immanuel - the doctor behind unproven coronavirus cure claim
(2020).
\url{https://www.bbc.com/news/world-africa-53579773}
\end{botherref}
\endbibitem

\bibitem{comm_tradingstandards}
\begin{botherref}
\oauthor{\bsnm{Institute}, \binits{C.T.S.}}:
New Covid-19 app exploited by fraudsters to scam public
(2020).
\url{https://www.tradingstandards.uk/news-policy/news-room/2020/new-covid-19-app-exploited-by-fraudsters-to-scam-public}
\end{botherref}
\endbibitem

\bibitem{Vijjali_transformer_covid_fake_news}
\begin{bchapter}
\bauthor{\bsnm{Vijjali}, \binits{R.}},
\bauthor{\bsnm{Potluri}, \binits{P.}},
\bauthor{\bsnm{Kumar}, \binits{S.}},
\bauthor{\bsnm{Sundeep}, \binits{T.}}:
\bctitle{Two stage transformer model for covid-19 fake news detection and fact
  checking}.
In: \bbtitle{Proceedings of the Workshop on NLP for Internet Freedom}
(\byear{2020})
\end{bchapter}
\endbibitem

\bibitem{brown_2009}
\begin{botherref}
\oauthor{\bsnm{Brown}, \binits{D.}}:
Most of any vaccine for new flu strain could be claimed by Rich Nations'
  preexisting contracts.
WP Company
(2009).
\url{https://www.washingtonpost.com/wp-dyn/content/article/2009/05/06/AR2009050603760.html}
\end{botherref}
\endbibitem

\bibitem{world_trade_oranization_2020}
\begin{botherref}
\oauthor{\bsnm{Oranization}, \binits{W.T.}}:
Export Prohibitions and Restrictions
(2020).
\url{https://www.wto.org/english/tratop_e/covid19_e/export_prohibitions_report_e.pdf}
\end{botherref}
\endbibitem

\bibitem{fawcett2021middle}
\begin{barticle}
\bauthor{\bsnm{Fawcett}, \binits{L.}}:
\batitle{The middle east and covid-19: time for collective action}.
\bjtitle{Globalization and Health}
\bvolume{17}(\bissue{1}),
\bfpage{1}--\blpage{9}
(\byear{2021})
\end{barticle}
\endbibitem

\bibitem{Delta_2021}
\begin{botherref}
\oauthor{\bsnm{Humanitarian}, \binits{T.N.}}:
(2021).
\url{https://www.thenewhumanitarian.org/news/2021/8/18/soaring-cases-and-little-vaccination-a-covid-19-middle-east-snapshot}
\end{botherref}
\endbibitem

\bibitem{Middle_East_Yemen}
\begin{botherref}
Middle East in 2021: Despite a year out of the global spotlight, millions
  remain in need
(2021).
\url{https://reliefweb.int/report/yemen/middle-east-2021-despite-year-out-global-spotlight-millions-remain-need}
\end{botherref}
\endbibitem

\bibitem{essar2021covid}
\begin{barticle}
\bauthor{\bsnm{Essar}, \binits{M.Y.}},
\bauthor{\bsnm{Hasan}, \binits{M.M.}},
\bauthor{\bsnm{Islam}, \binits{Z.}},
\bauthor{\bsnm{Riaz}, \binits{M.M.A.}},
\bauthor{\bsnm{Aborode}, \binits{A.T.}},
\bauthor{\bsnm{Ahmad}, \binits{S.}}:
\batitle{Covid-19 and multiple crises in afghanistan: an urgent battle}.
\bjtitle{Conflict and Health}
\bvolume{15}(\bissue{1}),
\bfpage{1}--\blpage{3}
(\byear{2021})
\end{barticle}
\endbibitem

\bibitem{Afghanistan_WHO_Portal}
\begin{botherref}
\oauthor{\bsnm{WHO}}:
COVID-19 WHO Portal for Afghanistan.
\url{https://covid19.who.int/region/emro/country/af/}
\end{botherref}
\endbibitem

\bibitem{COVID_Iraq_2021}
\begin{botherref}
\oauthor{\bsnm{Al-Saiedi}, \binits{A.}}:
Latest COVID-19 Surge Pushes More Iraqis to Get Vaccinated, But Hesitancy Still
  Remains
(2021).
\url{https://phr.org/our-work/resources/latest-covid-19-surge-push-more-iraqis-to-get-vaccinated-but-hesitancy-still-remains/}
\end{botherref}
\endbibitem

\bibitem{unicef_2021}
\begin{botherref}
\oauthor{\bsnm{(UNICEF)}, \binits{P.R.}}:
First Covid-19 COVAX vaccine doses administered in Africa
(2021).
\url{https://www.unicef.org/press-releases/first-covid-19-covax-vaccine-doses-administered-africa}
\end{botherref}
\endbibitem

\bibitem{COVID_vaccination_2021}
\begin{botherref}
\oauthor{\bsnm{CDC}, \binits{A.}}:
(2021).
\url{https://africacdc.org/covid-19-vaccination/}
\end{botherref}
\endbibitem

\bibitem{karafillakis2017benefit}
\begin{barticle}
\bauthor{\bsnm{Karafillakis}, \binits{E.}},
\bauthor{\bsnm{Larson}, \binits{H.J.}}, \betal:
\batitle{The benefit of the doubt or doubts over benefits? a systematic
  literature review of perceived risks of vaccines in european populations}.
\bjtitle{Vaccine}
\bvolume{35}(\bissue{37}),
\bfpage{4840}--\blpage{4850}
(\byear{2017})
\end{barticle}
\endbibitem

\bibitem{who_2021}
\begin{botherref}
\oauthor{\bsnm{Africa}, \binits{W.}}:
Key lessons from Africa's COVID-19 vaccine rollout.
World Health Organization
(2021).
\url{https://www.afro.who.int/news/key-lessons-africas-covid-19-vaccine-rollout}
\end{botherref}
\endbibitem

\bibitem{mwai_2021}
\begin{botherref}
\oauthor{\bsnm{Mwai}, \binits{P.}}:
Covid-19 vaccinations: African nations miss who target.
BBC
(2021).
\url{https://www.bbc.com/news/56100076}
\end{botherref}
\endbibitem

\bibitem{equity2021covid}
\begin{barticle}
\bauthor{\bsnm{Equity}, \binits{V.}}:
\batitle{Covid-19 vaccine equity and booster doses}.
\bjtitle{Lancet Infect Dis}
\bvolume{21},
\bfpage{743}
(\byear{2021})
\end{barticle}
\endbibitem

\bibitem{mutombo2021covid}
\begin{botherref}
\oauthor{\bsnm{Mutombo}, \binits{P.N.}},
\oauthor{\bsnm{Fallah}, \binits{M.P.}},
\oauthor{\bsnm{Munodawafa}, \binits{D.}},
\oauthor{\bsnm{Kabel}, \binits{A.}},
\oauthor{\bsnm{Houeto}, \binits{D.}},
\oauthor{\bsnm{Goronga}, \binits{T.}},
\oauthor{\bsnm{Mweemba}, \binits{O.}},
\oauthor{\bsnm{Balance}, \binits{G.}},
\oauthor{\bsnm{Onya}, \binits{H.}},
\oauthor{\bsnm{Kamba}, \binits{R.S.}}, et al.:
Covid-19 vaccine hesitancy in africa: a call to action.
The Lancet. Global Health
(2021)
\end{botherref}
\endbibitem

\bibitem{sallam2021covid}
\begin{barticle}
\bauthor{\bsnm{Sallam}, \binits{M.}}:
\batitle{Covid-19 vaccine hesitancy worldwide: a concise systematic review of
  vaccine acceptance rates}.
\bjtitle{Vaccines}
\bvolume{9}(\bissue{2}),
\bfpage{160}
(\byear{2021})
\end{barticle}
\endbibitem

\bibitem{jerving2021long}
\begin{barticle}
\bauthor{\bsnm{Jerving}, \binits{S.}}:
\batitle{The long road ahead for covid-19 vaccination in africa}.
\bjtitle{The Lancet}
\bvolume{398}(\bissue{10303}),
\bfpage{827}--\blpage{828}
(\byear{2021})
\end{barticle}
\endbibitem

\bibitem{Reuters_2021}
\begin{botherref}
\oauthor{\bsnm{Reuters}}:
Eritrea has not started vaccinating against covid, says africa cdc.
Reuters
(2021)
\end{botherref}
\endbibitem

\bibitem{makoni2020covid}
\begin{barticle}
\bauthor{\bsnm{Makoni}, \binits{M.}}:
\batitle{Covid-19 vaccine trials in africa}.
\bjtitle{The Lancet Respiratory Medicine}
\bvolume{8}(\bissue{11}),
\bfpage{79}--\blpage{80}
(\byear{2020})
\end{barticle}
\endbibitem

\bibitem{croda2021booster}
\begin{botherref}
\oauthor{\bsnm{Croda}, \binits{J.}},
\oauthor{\bsnm{Ranzani}, \binits{O.T.}}:
Booster doses for inactivated covid-19 vaccines: if, when, and for whom.
The Lancet Infectious Diseases
(2021)
\end{botherref}
\endbibitem

\bibitem{krause2021considerations}
\begin{barticle}
\bauthor{\bsnm{Krause}, \binits{P.R.}},
\bauthor{\bsnm{Fleming}, \binits{T.R.}},
\bauthor{\bsnm{Peto}, \binits{R.}},
\bauthor{\bsnm{Longini}, \binits{I.M.}},
\bauthor{\bsnm{Figueroa}, \binits{J.P.}},
\bauthor{\bsnm{Sterne}, \binits{J.A.}},
\bauthor{\bsnm{Cravioto}, \binits{A.}},
\bauthor{\bsnm{Rees}, \binits{H.}},
\bauthor{\bsnm{Higgins}, \binits{J.P.}},
\bauthor{\bsnm{Boutron}, \binits{I.}}, \betal:
\batitle{Considerations in boosting covid-19 vaccine immune responses}.
\bjtitle{The Lancet}
\bvolume{398}(\bissue{10308}),
\bfpage{1377}--\blpage{1380}
(\byear{2021})
\end{barticle}
\endbibitem

\bibitem{Lieber_2021}
\begin{botherref}
\oauthor{\bsnm{Lieber}, \binits{D.}}:
Israel begins pfizer booster shots for at-risk adults as delta cases rise.
Wall Street journal (Eastern ed.)
(2021)
\end{botherref}
\endbibitem

\bibitem{CDC_2022}
\begin{botherref}
\oauthor{\bsnm{CDC}}:
COVID-19 vaccine booster shots
(2022).
\url{https://www.cdc.gov/coronavirus/2019-ncov/vaccines/booster-shot.html}
\end{botherref}
\endbibitem

\bibitem{WHO_Director-General_2021}
\begin{botherref}
WHO director general's opening remarks at the media briefing on COVID 4 August
  2021.
WHO-Director General
(2021).
\url{https://www.who.int/director-general/speeches/detail/who-director-general-s-opening-remarks-at-the-media-briefing-on-covid-4-august-2021}
\end{botherref}
\endbibitem

\end{thebibliography}

\newcommand{\BMCxmlcomment}[1]{}

\BMCxmlcomment{

<refgrp>

<bibl id="B1">
  <title><p>We can’t defeat COVID-19 without vaccinating children. There
  aren't even any kids' clinical trials yet</p></title>
  <aug>
    <au><cnm>AAMC</cnm></au>
  </aug>
  <pubdate>2020</pubdate>
  <url>https://www.aamc.org/news-insights/we-can-t-defeat-covid-19-without-vaccinating-children-there-arent-even-any-kids-clinical-trials-yet</url>
</bibl>

<bibl id="B2">
  <title><p>Immunosenescence and human vaccine immune responses</p></title>
  <aug>
    <au><snm>Crooke</snm><fnm>SN</fnm></au>
    <au><snm>Ovsyannikova</snm><fnm>IG</fnm></au>
    <au><snm>Poland</snm><fnm>GA</fnm></au>
    <au><snm>Kennedy</snm><fnm>RB</fnm></au>
  </aug>
  <source>Immunity \& Ageing</source>
  <publisher>Springer</publisher>
  <pubdate>2019</pubdate>
  <volume>16</volume>
  <issue>1</issue>
  <fpage>25</fpage>
</bibl>

<bibl id="B3">
  <title><p>Draft landscape of covid 19 candidate vaccines</p></title>
  <aug>
    <au><cnm>WHO</cnm></au>
  </aug>
  <pubdate>2020</pubdate>
  <url>https://www.who.int/publications/m/item/draft-landscape-of-covid-19-candidate-vaccines</url>
</bibl>

<bibl id="B4">
  <title><p>Revealing public opinion towards COVID-19 vaccines using Twitter
  data in the United States: a spatiotemporal perspective</p></title>
  <aug>
    <au><snm>Hu</snm><fnm>T</fnm></au>
    <au><snm>Wang</snm><fnm>S</fnm></au>
    <au><snm>Luo</snm><fnm>W</fnm></au>
    <au><snm>Yan</snm><fnm>Y</fnm></au>
    <au><snm>Zhang</snm><fnm>M</fnm></au>
    <au><snm>Huang</snm><fnm>X</fnm></au>
    <au><snm>Liu</snm><fnm>R</fnm></au>
    <au><snm>Ly</snm><fnm>K</fnm></au>
    <au><snm>Kacker</snm><fnm>V</fnm></au>
    <au><snm>Li</snm><fnm>Z</fnm></au>
  </aug>
  <source>medRxiv</source>
  <publisher>Cold Spring Harbor Laboratory Press</publisher>
  <pubdate>2021</pubdate>
</bibl>

<bibl id="B5">
  <title><p>COVID-19 vaccine challenges: What have we learned so far and what
  remains to be done?</p></title>
  <aug>
    <au><snm>Forman</snm><fnm>R</fnm></au>
    <au><snm>Shah</snm><fnm>S</fnm></au>
    <au><snm>Jeurissen</snm><fnm>P</fnm></au>
    <au><snm>Jit</snm><fnm>M</fnm></au>
    <au><snm>Mossialos</snm><fnm>E</fnm></au>
  </aug>
  <source>Health Policy</source>
  <publisher>Elsevier</publisher>
  <pubdate>2021</pubdate>
</bibl>

<bibl id="B6">
  <title><p>Vaccine distribution—equity left behind?</p></title>
  <aug>
    <au><snm>Jean Jacques</snm><fnm>M</fnm></au>
    <au><snm>Bauchner</snm><fnm>H</fnm></au>
  </aug>
  <source>JAMA</source>
  <publisher>American Medical Association</publisher>
  <pubdate>2021</pubdate>
  <volume>325</volume>
  <issue>9</issue>
  <fpage>829</fpage>
  <lpage>-830</lpage>
</bibl>

<bibl id="B7">
  <title><p>Equal Access of COVID-19 Vaccine distribution in Africa: Challenges
  and Way Forward</p></title>
  <aug>
    <au><snm>Aborode</snm><fnm>AT</fnm></au>
    <au><snm>Olofinsao</snm><fnm>OA</fnm></au>
    <au><snm>Osmond</snm><fnm>E</fnm></au>
    <au><snm>Batubo</snm><fnm>AP</fnm></au>
    <au><snm>Fayemiro</snm><fnm>O</fnm></au>
    <au><snm>Sherifdeen</snm><fnm>O</fnm></au>
    <au><snm>Muraina</snm><fnm>L</fnm></au>
    <au><snm>Obadawo</snm><fnm>B</fnm></au>
    <au><snm>Ahmad</snm><fnm>S</fnm></au>
    <au><snm>Fajemisin</snm><fnm>EA</fnm></au>
  </aug>
  <source>Journal of Medical Virology</source>
  <publisher>Wiley Online Library</publisher>
  <pubdate>2021</pubdate>
</bibl>

<bibl id="B8">
  <title><p>A Review of the Progress and Challenges of Developing a Vaccine for
  COVID-19</p></title>
  <aug>
    <au><snm>Sharma</snm><fnm>O</fnm></au>
    <au><snm>Sultan</snm><fnm>AA</fnm></au>
    <au><snm>Ding</snm><fnm>H</fnm></au>
    <au><snm>Triggle</snm><fnm>CR</fnm></au>
  </aug>
  <source>Frontiers in immunology</source>
  <publisher>Frontiers</publisher>
  <pubdate>2020</pubdate>
  <volume>11</volume>
  <fpage>2413</fpage>
</bibl>

<bibl id="B9">
  <title><p>COVID-19 and risks to the supply and quality of tests, drugs, and
  vaccines</p></title>
  <aug>
    <au><snm>Newton</snm><fnm>PN</fnm></au>
    <au><snm>Bond</snm><fnm>KC</fnm></au>
    <au><snm>Adeyeye</snm><fnm>M</fnm></au>
    <au><snm>Antignac</snm><fnm>M</fnm></au>
    <au><snm>Ashenef</snm><fnm>A</fnm></au>
    <au><snm>Awab</snm><fnm>GR</fnm></au>
    <au><snm>Bannenberg</snm><fnm>WJ</fnm></au>
    <au><snm>Bower</snm><fnm>J</fnm></au>
    <au><snm>Breman</snm><fnm>J</fnm></au>
    <au><snm>Brock</snm><fnm>A</fnm></au>
    <au><cnm>others</cnm></au>
  </aug>
  <source>The Lancet Global Health</source>
  <publisher>Elsevier</publisher>
  <pubdate>2020</pubdate>
  <volume>8</volume>
  <issue>6</issue>
  <fpage>e754</fpage>
  <lpage>-e755</lpage>
</bibl>

<bibl id="B10">
  <title><p>Challenges to COVID-19 vaccine supply chain: Implications for
  sustainable development goals</p></title>
  <aug>
    <au><snm>Alam</snm><fnm>ST</fnm></au>
    <au><snm>Ahmed</snm><fnm>S</fnm></au>
    <au><snm>Ali</snm><fnm>SM</fnm></au>
    <au><snm>Sarker</snm><fnm>S</fnm></au>
    <au><snm>Kabir</snm><fnm>G</fnm></au>
    <au><cnm>others</cnm></au>
  </aug>
  <source>International Journal of Production Economics</source>
  <publisher>Elsevier</publisher>
  <pubdate>2021</pubdate>
  <volume>239</volume>
  <fpage>108193</fpage>
</bibl>

<bibl id="B11">
  <title><p>A Framework for Equitable Allocation of COVID-19
  Vaccine</p></title>
  <aug>
    <au><snm>Kahn</snm><fnm>B</fnm></au>
    <au><snm>Brown</snm><fnm>L</fnm></au>
    <au><snm>Foege</snm><fnm>W</fnm></au>
    <au><snm>Gayle</snm><fnm>H</fnm></au>
    <au><snm>Sciences Engineering</snm><fnm>NA</fnm></au>
    <au><cnm>Medicine</cnm></au>
    <au><cnm>others</cnm></au>
  </aug>
  <source>Framework for Equitable Allocation of COVID-19 Vaccine</source>
  <publisher>National Academies Press (US)</publisher>
  <pubdate>2020</pubdate>
</bibl>

<bibl id="B12">
  <title><p>Fair allocation mechanism for COVID-19 vaccines through the COVAX
  Facility</p></title>
  <aug>
    <au><cnm>WHO</cnm></au>
  </aug>
  <pubdate>2020</pubdate>
  <url>https://www.who.int/publications/m/item/fair-allocation-mechanism-for-covid-19-vaccines-through-the-covax-facility</url>
</bibl>

<bibl id="B13">
  <title><p>COVID-19 Vaccination Program Interim Playbook for jurisdiction
  Operations</p></title>
  <aug>
    <au><cnm>CDC</cnm></au>
  </aug>
  <pubdate>2020</pubdate>
  <url>https://www.cdc.gov/vaccines/imz-managers/downloads/COVID-19-Vaccination-Program-Interim_Playbook.pdf</url>
</bibl>

<bibl id="B14">
  <title><p>Novelty, conformity and trust in vaccines - full text
  view</p></title>
  <source>Full Text View - ClinicalTrials.gov</source>
  <url>https://clinicaltrials.gov/ct2/show/NCT04693689</url>
</bibl>

<bibl id="B15">
  <title><p>Explaining operation warp speed</p></title>
  <aug>
    <au><cnm>HHS</cnm></au>
  </aug>
  <pubdate>2020</pubdate>
  <url>https://www.hhs.gov/coronavirus/explaining-operation-warp-speed/index.html</url>
</bibl>

<bibl id="B16">
  <title><p>Covax Explained</p></title>
  <aug>
    <au><cnm>GAVI</cnm></au>
  </aug>
  <pubdate>2020</pubdate>
  <url>https://www.gavi.org/vaccineswork/covax-explained</url>
</bibl>

<bibl id="B17">
  <title><p>Moderna\'s covid-19 vaccine candidate meets its primary
  efficacy</p></title>
  <aug>
    <au><cnm>Moderna</cnm></au>
  </aug>
  <pubdate>2020</pubdate>
  <url>https://investors.modernatx.com/news-releases/news-release-details/modernas-covid-19-vaccine-candidate-meets-its-primary-efficacy</url>
</bibl>

<bibl id="B18">
  <title><p>pfizer and biontech conclude phase 3 study covid-19
  vaccine</p></title>
  <aug>
    <au><cnm>pfizer</cnm></au>
  </aug>
  <pubdate>2020</pubdate>
  <url>https://www.pfizer.com/news/press-release/press-release-detail/pfizer-and-biontech-conclude-phase-3-study-covid-19-vaccine</url>
</bibl>

<bibl id="B19">
  <title><p>THE BIDEN PLAN TO COMBAT CORONAVIRUS (COVID-19) AND PREPARE FOR
  FUTURE GLOBAL HEALTH THREATS</p></title>
  <aug>
    <au><snm>Biden</snm><fnm>J</fnm></au>
  </aug>
  <pubdate>2020</pubdate>
  <url>https://joebiden.com/covid-plan/</url>
</bibl>

<bibl id="B20">
  <title><p>JOE AND KAMALA'S PLAN TO BEAT COVID-19</p></title>
  <aug>
    <au><snm>Biden</snm><fnm>J</fnm></au>
  </aug>
  <pubdate>2020</pubdate>
  <url>https://joebiden.com/covid19/</url>
</bibl>

<bibl id="B21">
  <title><p>How the Flu Virus Can Change: “Drift” and
  “Shift”</p></title>
  <aug>
    <au><cnm>CDC</cnm></au>
  </aug>
  <pubdate>2020</pubdate>
  <url>https://www.cdc.gov/flu/about/viruses/change.htm</url>
</bibl>

<bibl id="B22">
  <title><p>Vaccination and antigenic drift in influenza</p></title>
  <aug>
    <au><snm>Boni</snm><fnm>MF</fnm></au>
  </aug>
  <source>Vaccine</source>
  <pubdate>2008</pubdate>
  <volume>26</volume>
</bibl>

<bibl id="B23">
  <title><p>Guidance from Pediatric Stakeholders: A Coordinated Approach to
  Communicating Pediatric-related Information on Pandemic Influenza at the
  Community Level</p></title>
  <url>https://www.cdc.gov/h1n1flu/guidance/pediatrics_tool.htm</url>
</bibl>

<bibl id="B24">
  <title><p>Pandemic preparedness and response—lessons from the H1N1
  influenza of 2009</p></title>
  <aug>
    <au><snm>Fineberg</snm><fnm>HV</fnm></au>
  </aug>
  <source>New England Journal of Medicine</source>
  <publisher>Mass Medical Soc</publisher>
  <pubdate>2014</pubdate>
  <volume>370</volume>
  <issue>14</issue>
  <fpage>1335</fpage>
  <lpage>-1342</lpage>
</bibl>

<bibl id="B25">
  <title><p>Lessons learned from the A (H1N1) influenza pandemic</p></title>
  <aug>
    <au><snm>Vousden</snm><fnm>N</fnm></au>
    <au><snm>Knight</snm><fnm>M</fnm></au>
  </aug>
  <source>Best Practice \& Research Clinical Obstetrics \& Gynaecology</source>
  <publisher>Elsevier</publisher>
  <pubdate>2020</pubdate>
</bibl>

<bibl id="B26">
  <title><p>COVID-19: Federal Efforts Accelerate Vaccine and Therapeutic
  Development, but More Transparency Needed on Emergency Use
  Authorizations</p></title>
  <aug>
    <au><cnm>{U.S. Government Accountability Office}</cnm></au>
  </aug>
  <pubdate>2020</pubdate>
  <url>https://www.gao.gov/products/GAO-21-207</url>
</bibl>

<bibl id="B27">
  <title><p>{ACIP} December 2020 Presentation Slides {\textbar} Immunization
  Practices {\textbar} {CDC}</p></title>
  <aug>
    <au><cnm>{CDC}</cnm></au>
  </aug>
  <url>https://www.cdc.gov/vaccines/acip/meetings/slides-2020-12.html</url>
</bibl>

<bibl id="B28">
  <title><p>National Governors Association Submits List of Questions to Trump
  Administration on Effective Implementation of COVID-19 Vaccine</p></title>
  <aug>
    <au><snm>State</snm><fnm>NY</fnm></au>
  </aug>
  <pubdate>2020</pubdate>
  <url>https://www.governor.ny.gov/news/national-governors-association-submits-list-questions-trump-administration-effective</url>
</bibl>

<bibl id="B29">
  <title><p>How to Ship a Vaccine at $\sim$80°C, and Other Obstacles in the
  Covid Fight</p></title>
  <aug>
    <au><snm>Times</snm><fnm>NY</fnm></au>
  </aug>
  <pubdate>2020</pubdate>
  <url>https://www.nytimes.com/2020/09/18/business/coronavirus-covid-vaccine-cold-frozen-logistics.html</url>
</bibl>

<bibl id="B30">
  <title><p>COVID-19: The logistic challenges of vaccine distribution in
  India</p></title>
  <aug>
    <au><snm>Herald</snm><fnm>N</fnm></au>
  </aug>
  <pubdate>2020</pubdate>
  <url>https://www.nationalheraldindia.com/india/covid-19-the-logistic-challenges-of-vaccine-distribution-in-india</url>
</bibl>

<bibl id="B31">
  <title><p>The biggest problem isn’t when a vaccine is ready. It’s how to
  distribute it</p></title>
  <aug>
    <au><cnm>Washingtonpost</cnm></au>
  </aug>
  <pubdate>2020</pubdate>
  <url>https://www.washingtonpost.com/outlook/2020/10/27/coronavirus-vaccine-distribution-plan/</url>
</bibl>

<bibl id="B32">
  <title><p>Improving immunization in Afghanistan: results from a
  cross-sectional community-based survey to assess routine immunization
  coverage</p></title>
  <aug>
    <au><snm>Mugali</snm><fnm>RR</fnm></au>
    <au><snm>Mansoor</snm><fnm>F</fnm></au>
    <au><snm>Parwiz</snm><fnm>S</fnm></au>
    <au><snm>Ahmad</snm><fnm>F</fnm></au>
    <au><snm>Safi</snm><fnm>N</fnm></au>
    <au><snm>Higgins Steele</snm><fnm>A</fnm></au>
    <au><snm>Varkey</snm><fnm>S</fnm></au>
  </aug>
  <source>BMC Public Health</source>
  <publisher>Springer</publisher>
  <pubdate>2017</pubdate>
  <volume>17</volume>
  <issue>1</issue>
  <fpage>290</fpage>
</bibl>

<bibl id="B33">
  <title><p>Country Engagement</p></title>
  <aug>
    <au><snm>Bank</snm><fnm>TW</fnm></au>
  </aug>
  <pubdate>2020</pubdate>
  <url>https://id4d.worldbank.org/country-engagement</url>
</bibl>

<bibl id="B34">
  <title><p>H1N1 vaccine was unevenly distributed across L.A. County, figures
  show</p></title>
  <aug>
    <au><snm>Times</snm><fnm>LA</fnm></au>
  </aug>
  <pubdate>2010</pubdate>
  <url>https://www.latimes.com/archives/la-xpm-2010-mar-01-la-me-h1n1-supply1-2010mar01-story.html</url>
</bibl>

<bibl id="B35">
  <title><p>Vaccination and GDP growth rates: Exploring the links in a
  conditional convergence framework</p></title>
  <aug>
    <au><snm>Masia</snm><fnm>NA</fnm></au>
    <au><snm>Smerling</snm><fnm>J</fnm></au>
    <au><snm>Kapfidze</snm><fnm>T</fnm></au>
    <au><snm>Manning</snm><fnm>R</fnm></au>
    <au><snm>Showalter</snm><fnm>M</fnm></au>
  </aug>
  <source>World Development</source>
  <publisher>Elsevier</publisher>
  <pubdate>2018</pubdate>
  <volume>103</volume>
  <fpage>88</fpage>
  <lpage>-99</lpage>
</bibl>

<bibl id="B36">
  <title><p>U.S. Public Health Response to the Measles Outbreak</p></title>
  <aug>
    <au><cnm>CDC</cnm></au>
  </aug>
  <pubdate>2019</pubdate>
  <url>https://www.cdc.gov/washington/testimony/2019/t20190227.htm</url>
</bibl>

<bibl id="B37">
  <title><p>Immune activation with HIV vaccines</p></title>
  <aug>
    <au><snm>Fauci</snm><fnm>AS</fnm></au>
    <au><snm>Marovich</snm><fnm>MA</fnm></au>
    <au><snm>Dieffenbach</snm><fnm>CW</fnm></au>
    <au><snm>Hunter</snm><fnm>E</fnm></au>
    <au><snm>Buchbinder</snm><fnm>SP</fnm></au>
  </aug>
  <source>Science</source>
  <publisher>American Association for the Advancement of Science</publisher>
  <pubdate>2014</pubdate>
  <volume>344</volume>
  <issue>6179</issue>
  <fpage>49</fpage>
  <lpage>-51</lpage>
</bibl>

<bibl id="B38">
  <title><p>Could {COVID}-19 {mRNA} vaccines cause autoimmune
  diseases?</p></title>
  <aug>
    <au><snm>White</snm><fnm>S</fnm></au>
  </aug>
  <source>BMJ</source>
  <pubdate>2020</pubdate>
  <url>https://www.bmj.com/content/371/bmj.m4347/rr-6</url>
</bibl>

<bibl id="B39">
  <title><p>Use of adenovirus type-5 vectored vaccines: a cautionary
  tale</p></title>
  <aug>
    <au><snm>Buchbinder</snm><fnm>SP</fnm></au>
    <au><snm>McElrath</snm><fnm>MJ</fnm></au>
    <au><snm>Dieffenbach</snm><fnm>C</fnm></au>
    <au><snm>Corey</snm><fnm>L</fnm></au>
  </aug>
  <source>The Lancet</source>
  <publisher>Elsevier</publisher>
  <pubdate>2020</pubdate>
</bibl>

<bibl id="B40">
  <title><p>New vaccine technologies to combat outbreak situations</p></title>
  <aug>
    <au><snm>Rauch</snm><fnm>S</fnm></au>
    <au><snm>Jasny</snm><fnm>E</fnm></au>
    <au><snm>Schmidt</snm><fnm>KE</fnm></au>
    <au><snm>Petsch</snm><fnm>B</fnm></au>
  </aug>
  <source>Frontiers in immunology</source>
  <publisher>Frontiers</publisher>
  <pubdate>2018</pubdate>
  <volume>9</volume>
  <fpage>1963</fpage>
</bibl>

<bibl id="B41">
  <title><p>mRNA vaccines—a new era in vaccinology</p></title>
  <aug>
    <au><snm>Pardi</snm><fnm>N</fnm></au>
    <au><snm>Hogan</snm><fnm>MJ</fnm></au>
    <au><snm>Porter</snm><fnm>FW</fnm></au>
    <au><snm>Weissman</snm><fnm>D</fnm></au>
  </aug>
  <source>Nature reviews Drug discovery</source>
  <publisher>Nature Publishing Group</publisher>
  <pubdate>2018</pubdate>
  <volume>17</volume>
  <issue>4</issue>
  <fpage>261</fpage>
</bibl>

<bibl id="B42">
  <title><p>Johnson &amp Johnson Prepares to Resume Phase 3 ENSEMBLE Trial of
  its Janssen COVID-19 Vaccine Candidate in the U.S.</p></title>
  <aug>
    <au><snm>Johnson</snm><fnm>J</fnm></au>
  </aug>
  <pubdate>2020</pubdate>
  <url>https://www.jnj.com/our-company/johnson-johnson-prepares-to-resume-phase-3-ensemble-trial-of-its-janssen-covid-19-vaccine-candidate-in-the-us</url>
</bibl>

<bibl id="B43">
  <title><p>COVID-19 vaccine AZD1222 clinical trials resumed in the
  UK</p></title>
  <aug>
    <au><cnm>Astrazeneca</cnm></au>
  </aug>
  <pubdate>2020</pubdate>
  <url>https://www.astrazeneca.com/media-centre/press-releases/2020/covid-19-vaccine-azd1222-clinical-trials-resumed-in-the-uk.html</url>
</bibl>

<bibl id="B44">
  <title><p>Strategies to overcome host immunity to adenovirus vectors in
  vaccine development</p></title>
  <aug>
    <au><snm>Thacker</snm><fnm>EE</fnm></au>
    <au><snm>Timares</snm><fnm>L</fnm></au>
    <au><snm>Matthews</snm><fnm>QL</fnm></au>
  </aug>
  <source>Expert review of vaccines</source>
  <publisher>Taylor \& Francis</publisher>
  <pubdate>2009</pubdate>
  <volume>8</volume>
  <issue>6</issue>
  <fpage>761</fpage>
  <lpage>-777</lpage>
</bibl>

<bibl id="B45">
  <title><p>Adenoviral vectors are the new COVID-19 vaccine front-runners. Can
  they overcome their checkered past?</p></title>
  <aug>
    <au><cnm>C\&EN</cnm></au>
  </aug>
  <pubdate>2020</pubdate>
  <url>https://cen.acs.org/pharmaceuticals/vaccines/Adenoviral-vectors-new-COVID-19/98/i19</url>
</bibl>

<bibl id="B46">
  <title><p>Multiple sclerosis disease-modifying therapy and the COVID-19
  pandemic: implications on the risk of infection and future
  vaccination</p></title>
  <aug>
    <au><snm>Zheng</snm><fnm>C</fnm></au>
    <au><snm>Kar</snm><fnm>I</fnm></au>
    <au><snm>Chen</snm><fnm>CK</fnm></au>
    <au><snm>Sau</snm><fnm>C</fnm></au>
    <au><snm>Woodson</snm><fnm>S</fnm></au>
    <au><snm>Serra</snm><fnm>A</fnm></au>
    <au><snm>Abboud</snm><fnm>H</fnm></au>
  </aug>
  <source>CNS drugs</source>
  <publisher>Springer</publisher>
  <pubdate>2020</pubdate>
  <volume>34</volume>
  <issue>9</issue>
  <fpage>879</fpage>
  <lpage>-896</lpage>
</bibl>

<bibl id="B47">
  <title><p>Seasonal coronavirus protective immunity is
  short-lasting</p></title>
  <aug>
    <au><snm>Edridge</snm><fnm>AW</fnm></au>
    <au><snm>Kaczorowska</snm><fnm>J</fnm></au>
    <au><snm>Hoste</snm><fnm>AC</fnm></au>
    <au><snm>Bakker</snm><fnm>M</fnm></au>
    <au><snm>Klein</snm><fnm>M</fnm></au>
    <au><snm>Loens</snm><fnm>K</fnm></au>
    <au><snm>Jebbink</snm><fnm>MF</fnm></au>
    <au><snm>Matser</snm><fnm>A</fnm></au>
    <au><snm>Kinsella</snm><fnm>CM</fnm></au>
    <au><snm>Rueda</snm><fnm>P</fnm></au>
    <au><cnm>others</cnm></au>
  </aug>
  <source>Nature medicine</source>
  <publisher>Nature Publishing Group</publisher>
  <pubdate>2020</pubdate>
  <fpage>1</fpage>
  <lpage>-3</lpage>
</bibl>

<bibl id="B48">
  <title><p>Immunological memory to SARS-CoV-2 assessed for greater than six
  months after infection</p></title>
  <aug>
    <au><snm>Dan</snm><fnm>JM</fnm></au>
    <au><snm>Mateus</snm><fnm>J</fnm></au>
    <au><snm>Kato</snm><fnm>Y</fnm></au>
    <au><snm>Hastie</snm><fnm>KM</fnm></au>
    <au><snm>Faliti</snm><fnm>C</fnm></au>
    <au><snm>Ramirez</snm><fnm>SI</fnm></au>
    <au><snm>Frazier</snm><fnm>A</fnm></au>
    <au><snm>Esther</snm><fnm>DY</fnm></au>
    <au><snm>Grifoni</snm><fnm>A</fnm></au>
    <au><snm>Rawlings</snm><fnm>SA</fnm></au>
    <au><cnm>others</cnm></au>
  </aug>
  <source>bioRxiv</source>
  <publisher>Cold Spring Harbor Laboratory</publisher>
  <pubdate>2020</pubdate>
</bibl>

<bibl id="B49">
  <title><p>Duration of Effective Antibody Levels After COVID-19</p></title>
  <aug>
    <au><snm>Cruz</snm><fnm>AT</fnm></au>
    <au><snm>Zeichner</snm><fnm>SL</fnm></au>
  </aug>
  <source>Pediatrics</source>
  <publisher>American Academy of Pediatrics</publisher>
  <pubdate>2021</pubdate>
  <volume>148</volume>
  <issue>3</issue>
</bibl>

<bibl id="B50">
  <title><p>Seven-month kinetics of SARS-CoV-2 antibodies and role of
  pre-existing antibodies to human coronaviruses</p></title>
  <aug>
    <au><snm>Ortega</snm><fnm>N</fnm></au>
    <au><snm>Ribes</snm><fnm>M</fnm></au>
    <au><snm>Vidal</snm><fnm>M</fnm></au>
    <au><snm>Rubio</snm><fnm>R</fnm></au>
    <au><snm>Aguilar</snm><fnm>R</fnm></au>
    <au><snm>Williams</snm><fnm>S</fnm></au>
    <au><snm>Barrios</snm><fnm>D</fnm></au>
    <au><snm>Alonso</snm><fnm>S</fnm></au>
    <au><snm>Hern{\'a}ndez Luis</snm><fnm>P</fnm></au>
    <au><snm>Mitchell</snm><fnm>RA</fnm></au>
    <au><cnm>others</cnm></au>
  </aug>
  <source>Nature Communications</source>
  <publisher>Nature Publishing Group</publisher>
  <pubdate>2021</pubdate>
  <volume>12</volume>
  <issue>1</issue>
  <fpage>1</fpage>
  <lpage>-10</lpage>
</bibl>

<bibl id="B51">
  <title><p>Duration of immunity against pertussis after natural infection or
  vaccination</p></title>
  <aug>
    <au><snm>Wendelboe</snm><fnm>AM</fnm></au>
    <au><snm>Van Rie</snm><fnm>A</fnm></au>
    <au><snm>Salmaso</snm><fnm>S</fnm></au>
    <au><snm>Englund</snm><fnm>JA</fnm></au>
  </aug>
  <source>The Pediatric infectious disease journal</source>
  <publisher>LWW</publisher>
  <pubdate>2005</pubdate>
  <volume>24</volume>
  <issue>5</issue>
  <fpage>S58</fpage>
  <lpage>-S61</lpage>
</bibl>

<bibl id="B52">
  <title><p>Optimizing the impact of low-efficacy influenza
  vaccines</p></title>
  <aug>
    <au><snm>Sah</snm><fnm>P</fnm></au>
    <au><snm>Medlock</snm><fnm>J</fnm></au>
    <au><snm>Fitzpatrick</snm><fnm>MC</fnm></au>
    <au><snm>Singer</snm><fnm>BH</fnm></au>
    <au><snm>Galvani</snm><fnm>AP</fnm></au>
  </aug>
  <source>Proceedings of the National Academy of Sciences</source>
  <publisher>National Acad Sciences</publisher>
  <pubdate>2018</pubdate>
  <volume>115</volume>
  <issue>20</issue>
  <fpage>5151</fpage>
  <lpage>-5156</lpage>
</bibl>

<bibl id="B53">
  <title><p>Decline in seasonal influenza vaccine effectiveness with
  vaccination program maturation: A systematic review and
  meta-analysis</p></title>
  <aug>
    <au><snm>Okoli</snm><fnm>GN</fnm></au>
    <au><snm>Racovitan</snm><fnm>F</fnm></au>
    <au><snm>Abdulwahid</snm><fnm>T</fnm></au>
    <au><snm>Hyder</snm><fnm>SK</fnm></au>
    <au><snm>Lansbury</snm><fnm>L</fnm></au>
    <au><snm>Righolt</snm><fnm>CH</fnm></au>
    <au><snm>Mahmud</snm><fnm>SM</fnm></au>
    <au><snm>Nguyen Van Tam</snm><fnm>JS</fnm></au>
  </aug>
  <source>Open forum infectious diseases</source>
  <pubdate>2021</pubdate>
  <volume>8</volume>
  <fpage>ofab069</fpage>
</bibl>

<bibl id="B54">
  <title><p>OUR PROGRESS IN DEVELOPING A POTENTIAL COVID-19 VACCINE</p></title>
  <aug>
    <au><cnm>Pfizer</cnm></au>
  </aug>
  <pubdate>2020</pubdate>
  <url>https://www.pfizer.com/science/coronavirus/vaccine</url>
</bibl>

<bibl id="B55">
  <title><p>The Exclusion of Older Persons From Vaccine and Treatment Trials
  for Coronavirus Disease 2019—Missing the Target</p></title>
  <aug>
    <au><snm>Helfand</snm><fnm>BK</fnm></au>
    <au><snm>Webb</snm><fnm>M</fnm></au>
    <au><snm>Gartaganis</snm><fnm>SL</fnm></au>
    <au><snm>Fuller</snm><fnm>L</fnm></au>
    <au><snm>Kwon</snm><fnm>CS</fnm></au>
    <au><snm>Inouye</snm><fnm>SK</fnm></au>
  </aug>
  <source>JAMA Internal Medicine</source>
  <pubdate>2020</pubdate>
</bibl>

<bibl id="B56">
  <title><p>Development and Licensure of Vaccines to Prevent COVID-19 Guidance
  for Industry</p></title>
  <aug>
    <au><cnm>FDA</cnm></au>
  </aug>
  <pubdate>2020</pubdate>
  <url>https://www.fda.gov/media/139638/download</url>
</bibl>

<bibl id="B57">
  <title><p>The Americans with disabilities act and healthcare
  employer-mandated vaccinations</p></title>
  <aug>
    <au><snm>Yang</snm><fnm>YT</fnm></au>
    <au><snm>Pendo</snm><fnm>E</fnm></au>
    <au><snm>Reiss</snm><fnm>DR</fnm></au>
  </aug>
  <source>Vaccine</source>
  <pubdate>2020</pubdate>
  <volume>38</volume>
  <issue>16</issue>
  <fpage>3184</fpage>
  <lpage>3186</lpage>
  <url>http://www.sciencedirect.com/science/article/pii/S0264410X2030356X</url>
</bibl>

<bibl id="B58">
  <title><p>U.S. Public Now Divided Over Whether To Get COVID-19
  Vaccine</p></title>
  <aug>
    <au><snm>Research</snm><fnm>P</fnm></au>
  </aug>
  <pubdate>2020</pubdate>
  <url>https://www.pewresearch.org/science/2020/09/17/u-s-public-now-divided-over-whether-to-get-covid-19-vaccine/</url>
</bibl>

<bibl id="B59">
  <title><p>STATE OF THE UNION</p></title>
  <aug>
    <au><cnm>CNN</cnm></au>
  </aug>
  <pubdate>2020</pubdate>
  <url>http://transcripts.cnn.com/TRANSCRIPTS/2011/15/sotu.01.html</url>
</bibl>

<bibl id="B60">
  <title><p>Frequently Asked Questions about COVID-19 Vaccination</p></title>
  <aug>
    <au><cnm>CDC</cnm></au>
  </aug>
  <pubdate>2020</pubdate>
  <url>https://www.cdc.gov/coronavirus/2019-ncov/vaccines/faq.html</url>
</bibl>

<bibl id="B61">
  <title><p>A global survey of potential acceptance of a COVID-19
  vaccine</p></title>
  <aug>
    <au><snm>Lazarus</snm><fnm>JV</fnm></au>
    <au><snm>Ratzan</snm><fnm>SC</fnm></au>
    <au><snm>Palayew</snm><fnm>A</fnm></au>
    <au><snm>Gostin</snm><fnm>LO</fnm></au>
    <au><snm>Larson</snm><fnm>HJ</fnm></au>
    <au><snm>Rabin</snm><fnm>K</fnm></au>
    <au><snm>Kimball</snm><fnm>S</fnm></au>
    <au><snm>El Mohandes</snm><fnm>A</fnm></au>
  </aug>
  <source>Nature medicine</source>
  <publisher>Nature Publishing Group</publisher>
  <pubdate>2020</pubdate>
  <fpage>1</fpage>
  <lpage>-4</lpage>
</bibl>

<bibl id="B62">
  <title><p>Draft landscape of COVID-19 candidate vaccines</p></title>
  <aug>
    <au><cnm>WHO</cnm></au>
  </aug>
  <pubdate>2020</pubdate>
  <url>https://www.who.int/publications/m/item/draft-landscape-of-covid-19-candidate-vaccines</url>
</bibl>

<bibl id="B63">
  <title><p>Safety and immunogenicity of two RNA-based Covid-19 vaccine
  candidates</p></title>
  <aug>
    <au><snm>Walsh</snm><fnm>EE</fnm></au>
    <au><snm>Frenck Jr</snm><fnm>RW</fnm></au>
    <au><snm>Falsey</snm><fnm>AR</fnm></au>
    <au><snm>Kitchin</snm><fnm>N</fnm></au>
    <au><snm>Absalon</snm><fnm>J</fnm></au>
    <au><snm>Gurtman</snm><fnm>A</fnm></au>
    <au><snm>Lockhart</snm><fnm>S</fnm></au>
    <au><snm>Neuzil</snm><fnm>K</fnm></au>
    <au><snm>Mulligan</snm><fnm>MJ</fnm></au>
    <au><snm>Bailey</snm><fnm>R</fnm></au>
    <au><cnm>others</cnm></au>
  </aug>
  <source>New England Journal of Medicine</source>
  <publisher>Mass Medical Soc</publisher>
  <pubdate>2020</pubdate>
</bibl>

<bibl id="B64">
  <title><p>Evaluation of the immunogenicity of prime-boost vaccination with
  the replication-deficient viral vectored COVID-19 vaccine candidate ChAdOx1
  nCoV-19</p></title>
  <aug>
    <au><snm>Graham</snm><fnm>SP</fnm></au>
    <au><snm>McLean</snm><fnm>RK</fnm></au>
    <au><snm>Spencer</snm><fnm>AJ</fnm></au>
    <au><snm>Belij Rammerstorfer</snm><fnm>S</fnm></au>
    <au><snm>Wright</snm><fnm>D</fnm></au>
    <au><snm>Ulaszewska</snm><fnm>M</fnm></au>
    <au><snm>Edwards</snm><fnm>JC</fnm></au>
    <au><snm>Hayes</snm><fnm>JW</fnm></au>
    <au><snm>Martini</snm><fnm>V</fnm></au>
    <au><snm>Thakur</snm><fnm>N</fnm></au>
    <au><cnm>others</cnm></au>
  </aug>
  <source>NPJ vaccines</source>
  <publisher>Nature Publishing Group</publisher>
  <pubdate>2020</pubdate>
  <volume>5</volume>
  <issue>1</issue>
  <fpage>1</fpage>
  <lpage>-6</lpage>
</bibl>

<bibl id="B65">
  <title><p>PanVax Tool for Pandemic Vaccination Planning</p></title>
  <aug>
    <au><cnm>CDC</cnm></au>
  </aug>
  <pubdate>2020</pubdate>
  <url>https://www.cdc.gov/flu/pandemic-resources/tools/panvax-tool.htm</url>
</bibl>

<bibl id="B66">
  <title><p>Vaccine Tracking System (VTrckS)</p></title>
  <aug>
    <au><cnm>CDC</cnm></au>
  </aug>
  <pubdate>2020</pubdate>
  <url>https://www.cdc.gov/vaccines/programs/vtrcks/index.html</url>
</bibl>

<bibl id="B67">
  <title><p>Palantir to Help U.S. Track Covid-19 Vaccines</p></title>
  <aug>
    <au><snm>Journal</snm><fnm>TWS</fnm></au>
  </aug>
  <pubdate>2020</pubdate>
  <url>https://www.wsj.com/articles/palantir-to-help-u-s-track-covid-19-vaccines-11603367276</url>
</bibl>

<bibl id="B68">
  <title><p>Vaccine Collaboration Hub from SAP Improves Supply Chain Efficiency
  for Government and Life Sciences Organizations</p></title>
  <aug>
    <au><cnm>SAP</cnm></au>
  </aug>
  <pubdate>2020</pubdate>
  <url>https://news.sap.com/2020/10/vaccine-collaboration-hub-supply-chain-efficiency-government-life-sciences/</url>
</bibl>

<bibl id="B69">
  <title><p>Accenture Vaccine Management Solution</p></title>
  <aug>
    <au><cnm>Accenture</cnm></au>
  </aug>
  <pubdate>2020</pubdate>
  <url>https://www.accenture.com/us-en/services/public-service/vaccine-management-solution</url>
</bibl>

<bibl id="B70">
  <title><p>COVID-19 Vaccination Plan</p></title>
  <aug>
    <au><snm>Health</snm><fnm>MD</fnm></au>
  </aug>
  <pubdate>2020</pubdate>
  <url>https://phpa.health.maryland.gov/Documents/10.19.2020_Maryland_COVID-19_Vaccination_Plan_CDCwm.pdf</url>
</bibl>

<bibl id="B71">
  <title><p>Want herd immunity? Pay people to take the vaccine</p></title>
  <aug>
    <au><cnm>Bookings</cnm></au>
  </aug>
  <pubdate>2020</pubdate>
  <url>https://www.brookings.edu/opinions/want-herd-immunity-pay-people-to-take-the-vaccine/</url>
</bibl>

<bibl id="B72">
  <title><p>ENACT: Encounter-Based Architecture for Contact Tracing</p></title>
  <aug>
    <au><snm>Prasad</snm><fnm>A</fnm></au>
    <au><snm>Kotz</snm><fnm>D</fnm></au>
  </aug>
  <source>Proceedings of the 4th International on Workshop on Physical
  Analytics</source>
  <publisher>New York, NY, USA: Association for Computing Machinery</publisher>
  <series><title><p>WPA '17</p></title></series>
  <pubdate>2017</pubdate>
  <fpage>37–42</fpage>
  <url>https://doi.org/10.1145/3092305.3092310</url>
</bibl>

<bibl id="B73">
  <title><p>Epione: Lightweight Contact Tracing with Strong Privacy</p></title>
  <aug>
    <au><snm>Trieu</snm><fnm>N</fnm></au>
    <au><snm>Shehata</snm><fnm>K</fnm></au>
    <au><snm>Saxena</snm><fnm>P</fnm></au>
    <au><snm>Shokri</snm><fnm>R</fnm></au>
    <au><snm>Song</snm><fnm>D</fnm></au>
  </aug>
  <pubdate>2020</pubdate>
</bibl>

<bibl id="B74">
  <title><p>Apps Gone Rogue: Maintaining Personal Privacy in an
  Epidemic</p></title>
  <aug>
    <au><snm>Raskar</snm><fnm>R</fnm></au>
    <au><snm>Schunemann</snm><fnm>I</fnm></au>
    <au><snm>Barbar</snm><fnm>R</fnm></au>
    <au><snm>Vilcans</snm><fnm>K</fnm></au>
    <au><snm>Gray</snm><fnm>J</fnm></au>
    <au><snm>Vepakomma</snm><fnm>P</fnm></au>
    <au><snm>Kapa</snm><fnm>S</fnm></au>
    <au><snm>Nuzzo</snm><fnm>A</fnm></au>
    <au><snm>Gupta</snm><fnm>R</fnm></au>
    <au><snm>Berke</snm><fnm>A</fnm></au>
    <au><snm>Greenwood</snm><fnm>D</fnm></au>
    <au><snm>Keegan</snm><fnm>C</fnm></au>
    <au><snm>Kanaparti</snm><fnm>S</fnm></au>
    <au><snm>Beaudry</snm><fnm>R</fnm></au>
    <au><snm>Stansbury</snm><fnm>D</fnm></au>
    <au><snm>Arcila</snm><fnm>BB</fnm></au>
    <au><snm>Kanaparti</snm><fnm>R</fnm></au>
    <au><snm>Pamplona</snm><fnm>V</fnm></au>
    <au><snm>Benedetti</snm><fnm>FM</fnm></au>
    <au><snm>Clough</snm><fnm>A</fnm></au>
    <au><snm>Das</snm><fnm>R</fnm></au>
    <au><snm>Jain</snm><fnm>K</fnm></au>
    <au><snm>Louisy</snm><fnm>K</fnm></au>
    <au><snm>Nadeau</snm><fnm>G</fnm></au>
    <au><snm>Pamplona</snm><fnm>V</fnm></au>
    <au><snm>Penrod</snm><fnm>S</fnm></au>
    <au><snm>Rajaee</snm><fnm>Y</fnm></au>
    <au><snm>Singh</snm><fnm>A</fnm></au>
    <au><snm>Storm</snm><fnm>G</fnm></au>
    <au><snm>Werner</snm><fnm>J</fnm></au>
  </aug>
  <pubdate>2020</pubdate>
</bibl>

<bibl id="B75">
  <title><p>Assessing Disease Exposure Risk with Location Data: A Proposal for
  Cryptographic Preservation of Privacy</p></title>
  <aug>
    <au><snm>Berke</snm><fnm>A</fnm></au>
    <au><snm>Bakker</snm><fnm>M</fnm></au>
    <au><snm>Vepakomma</snm><fnm>P</fnm></au>
    <au><snm>Larson</snm><fnm>K</fnm></au>
    <au><snm>Pentland</snm><fnm>AS</fnm></au>
  </aug>
  <pubdate>2020</pubdate>
</bibl>

<bibl id="B76">
  <title><p>TraceSecure: Towards Privacy Preserving Contact Tracing</p></title>
  <aug>
    <au><snm>Bell</snm><fnm>J</fnm></au>
    <au><snm>Butler</snm><fnm>D</fnm></au>
    <au><snm>Hicks</snm><fnm>C</fnm></au>
    <au><snm>Crowcroft</snm><fnm>J</fnm></au>
  </aug>
  <pubdate>2020</pubdate>
</bibl>

<bibl id="B77">
  <title><p>Political interference in public health science during
  covid-19</p></title>
  <aug>
    <au><snm>BJM</snm><fnm>T</fnm></au>
  </aug>
  <pubdate>2020</pubdate>
  <url>https://www.bmj.com/content/371/bmj.m3878</url>
</bibl>

<bibl id="B78">
  <title><p>Political interference in public health science during
  covid-19</p></title>
  <aug>
    <au><snm>BJM</snm><fnm>T</fnm></au>
  </aug>
  <pubdate>2020</pubdate>
  <url>https://www.bmj.com/content/371/bmj.m3878/rapid-responses</url>
</bibl>

<bibl id="B79">
  <title><p>Hydroxychloroquine promoted by Trump as Covid 'game changer' linked
  to increased deaths</p></title>
  <aug>
    <au><snm>Healthworld</snm><fnm>ET</fnm></au>
  </aug>
  <pubdate>2020</pubdate>
  <url>https://health.economictimes.indiatimes.com/news/diagnostics/hydroxychloroquine-promoted-by-trump-as-covid-game-changer-linked-to-increased-deaths/75785673</url>
</bibl>

<bibl id="B80">
  <title><p>Dr. Fauci says all the ‘valid’ scientific data shows
  hydroxychloroquine isn’t effective in treating coronavirus</p></title>
  <aug>
    <au><cnm>CNBC</cnm></au>
  </aug>
  <pubdate>2020</pubdate>
  <url>https://www.cnbc.com/2020/07/29/dr-fauci-says-all-the-valid-scientific-data-shows-hydroxychloroquine-isnt-effective-in-treating-coronavirus.html</url>
</bibl>

<bibl id="B81">
  <title><p>More than 150 countries engaged in COVID-19 vaccine global access
  facility</p></title>
  <aug>
    <au><cnm>WHO</cnm></au>
  </aug>
  <pubdate>2020</pubdate>
  <url>https://www.who.int/news/item/15-07-2020-more-than-150-countries-engaged-in-covid-19-vaccine-global-access-facility</url>
</bibl>

<bibl id="B82">
  <title><p>STAT-Harris Poll: The share of Americans interested in getting
  Covid-19 vaccine as soon as possible is dropping</p></title>
  <aug>
    <au><snm>News</snm><fnm>STAT</fnm></au>
  </aug>
  <pubdate>2020</pubdate>
  <url>https://www.statnews.com/pharmalot/2020/10/19/covid19-coronavirus-pandemic-vaccine-racial-disparities/</url>
</bibl>

<bibl id="B83">
  <title><p>Public Trust in Science with Dr. Anthony Fauci</p></title>
  <aug>
    <au><snm>Center</snm><fnm>TH</fnm></au>
  </aug>
  <pubdate>2020</pubdate>
  <url>https://www.youtube.com/watch?v=Az1kD5xnzS4&ab_channel=TheHastingsCenter</url>
</bibl>

<bibl id="B84">
  <title><p>Who will accept a COVID-19 vaccine?</p></title>
  <aug>
    <au><cnm>CIDRAP</cnm></au>
  </aug>
  <pubdate>2020</pubdate>
  <url>https://www.cidrap.umn.edu/news-perspective/2020/10/who-will-accept-covid-19-vaccine</url>
</bibl>

<bibl id="B85">
  <title><p>Health experts want to prioritize people of color for a Covid-19
  vaccine. But how should it be done?</p></title>
  <aug>
    <au><snm>News</snm><fnm>STAT</fnm></au>
  </aug>
  <pubdate>2020</pubdate>
  <url>https://www.statnews.com/2020/11/09/health-experts-want-to-prioritize-people-of-color-for-covid19-vaccine-but-how-should-it-be-done/?utm_sourceSTAT+Newsletters&utm_campaign=00a79720d8-Daily_Recap&utm_medium=email&utm_term=0_8cab1d7961-00a79720d8-152583781</url>
</bibl>

<bibl id="B86">
  <title><p>Is It Lawful and Ethical to Prioritize Racial Minorities for
  COVID-19 Vaccines?</p></title>
  <aug>
    <au><snm>Schmidt</snm><fnm>H</fnm></au>
    <au><snm>Gostin</snm><fnm>LO</fnm></au>
    <au><snm>Williams</snm><fnm>MA</fnm></au>
  </aug>
  <source>JAMA</source>
  <pubdate>2020</pubdate>
</bibl>

<bibl id="B87">
  <title><p>Fighting an Infodemic: COVID-19 Fake News Dataset</p></title>
  <aug>
    <au><snm>Patwa</snm><fnm>P</fnm></au>
    <au><snm>Sharma</snm><fnm>S</fnm></au>
    <au><snm>PYKL</snm><fnm>S</fnm></au>
    <au><snm>Guptha</snm><fnm>V</fnm></au>
    <au><snm>Kumari</snm><fnm>G</fnm></au>
    <au><snm>Akhtar</snm><fnm>MS</fnm></au>
    <au><snm>Ekbal</snm><fnm>A</fnm></au>
    <au><snm>Das</snm><fnm>A</fnm></au>
    <au><snm>Chakraborty</snm><fnm>T</fnm></au>
  </aug>
  <pubdate>2020</pubdate>
</bibl>

<bibl id="B88">
  <title><p>Hundreds die of poisoning in Iran as fake news suggests methanol
  cure for virus</p></title>
  <aug>
    <au><snm>Israel</snm><fnm>T</fnm></au>
  </aug>
  <pubdate>2020</pubdate>
  <url>https://www.timesofisrael.com/hundreds-die-of-poisoning-in-iran-as-fake-news-suggests-methanol-cure-for-virus/</url>
</bibl>

<bibl id="B89">
  <title><p>Coronavirus: Bill Gates ‘microchip’ conspiracy theory and other
  vaccine claims fact-checked</p></title>
  <aug>
    <au><cnm>BBC</cnm></au>
  </aug>
  <pubdate>2020</pubdate>
  <url>https://www.bbc.com/news/52847648</url>
</bibl>

<bibl id="B90">
  <title><p>Vaccine misinformation and social media</p></title>
  <aug>
    <au><snm>Burki</snm><fnm>T</fnm></au>
  </aug>
  <source>The Lancet Digital Health</source>
  <publisher>Elsevier</publisher>
  <pubdate>2019</pubdate>
  <volume>1</volume>
  <issue>6</issue>
  <fpage>e258</fpage>
  <lpage>-e259</lpage>
</bibl>

<bibl id="B91">
  <title><p>Stella Immanuel - the doctor behind unproven coronavirus cure
  claim</p></title>
  <aug>
    <au><cnm>BBC</cnm></au>
  </aug>
  <pubdate>2020</pubdate>
  <url>https://www.bbc.com/news/world-africa-53579773</url>
</bibl>

<bibl id="B92">
  <title><p>New Covid-19 app exploited by fraudsters to scam public</p></title>
  <aug>
    <au><snm>Institute</snm><fnm>CTS</fnm></au>
  </aug>
  <pubdate>2020</pubdate>
  <url>https://www.tradingstandards.uk/news-policy/news-room/2020/new-covid-19-app-exploited-by-fraudsters-to-scam-public</url>
</bibl>

<bibl id="B93">
  <title><p>Two Stage Transformer Model for COVID-19 Fake News Detection and
  Fact Checking</p></title>
  <aug>
    <au><snm>Vijjali</snm><fnm>R</fnm></au>
    <au><snm>Potluri</snm><fnm>P</fnm></au>
    <au><snm>Kumar</snm><fnm>S</fnm></au>
    <au><snm>Sundeep</snm><fnm>T</fnm></au>
  </aug>
  <source>Proceedings of the Workshop on NLP for Internet Freedom</source>
  <pubdate>2020</pubdate>
</bibl>

<bibl id="B94">
  <title><p>Most of any vaccine for new flu strain could be claimed by Rich
  Nations' preexisting contracts</p></title>
  <aug>
    <au><snm>Brown</snm><fnm>D</fnm></au>
  </aug>
  <source>The Washington Post</source>
  <publisher>WP Company</publisher>
  <pubdate>2009</pubdate>
  <url>https://www.washingtonpost.com/wp-dyn/content/article/2009/05/06/AR2009050603760.html</url>
</bibl>

<bibl id="B95">
  <title><p>Export Prohibitions and Restrictions</p></title>
  <aug>
    <au><snm>Oranization</snm><fnm>WT</fnm></au>
  </aug>
  <source>World Trade Organization</source>
  <pubdate>2020</pubdate>
  <url>https://www.wto.org/english/tratop_e/covid19_e/export_prohibitions_report_e.pdf</url>
</bibl>

<bibl id="B96">
  <title><p>The Middle East and COVID-19: time for collective
  action</p></title>
  <aug>
    <au><snm>Fawcett</snm><fnm>L</fnm></au>
  </aug>
  <source>Globalization and Health</source>
  <publisher>Springer</publisher>
  <pubdate>2021</pubdate>
  <volume>17</volume>
  <issue>1</issue>
  <fpage>1</fpage>
  <lpage>-9</lpage>
</bibl>

<bibl id="B97">
  <aug>
    <au><snm>Humanitarian</snm><fnm>TN</fnm></au>
  </aug>
  <source>The New Humanitarian</source>
  <pubdate>2021</pubdate>
  <url>https://www.thenewhumanitarian.org/news/2021/8/18/soaring-cases-and-little-vaccination-a-covid-19-middle-east-snapshot</url>
</bibl>

<bibl id="B98">
  <title><p>Middle East in 2021: Despite a year out of the global spotlight,
  millions remain in need</p></title>
  <source>ReliefWeb</source>
  <pubdate>2021</pubdate>
  <url>https://reliefweb.int/report/yemen/middle-east-2021-despite-year-out-global-spotlight-millions-remain-need</url>
</bibl>

<bibl id="B99">
  <title><p>COVID-19 and multiple crises in Afghanistan: an urgent
  battle</p></title>
  <aug>
    <au><snm>Essar</snm><fnm>MY</fnm></au>
    <au><snm>Hasan</snm><fnm>MM</fnm></au>
    <au><snm>Islam</snm><fnm>Z</fnm></au>
    <au><snm>Riaz</snm><fnm>MMA</fnm></au>
    <au><snm>Aborode</snm><fnm>AT</fnm></au>
    <au><snm>Ahmad</snm><fnm>S</fnm></au>
  </aug>
  <source>Conflict and Health</source>
  <publisher>Springer</publisher>
  <pubdate>2021</pubdate>
  <volume>15</volume>
  <issue>1</issue>
  <fpage>1</fpage>
  <lpage>-3</lpage>
</bibl>

<bibl id="B100">
  <title><p>COVID-19 WHO Portal for Afghanistan</p></title>
  <aug>
    <au><cnm>WHO</cnm></au>
  </aug>
  <source>Who.int</source>
  <url>https://covid19.who.int/region/emro/country/af/</url>
</bibl>

<bibl id="B101">
  <title><p>Latest COVID-19 Surge Pushes More Iraqis to Get Vaccinated, But
  Hesitancy Still Remains</p></title>
  <aug>
    <au><snm>Al Saiedi</snm><fnm>A</fnm></au>
  </aug>
  <source>Physicians for Human Rights</source>
  <pubdate>2021</pubdate>
  <url>https://phr.org/our-work/resources/latest-covid-19-surge-push-more-iraqis-to-get-vaccinated-but-hesitancy-still-remains/</url>
</bibl>

<bibl id="B102">
  <title><p>First Covid-19 COVAX vaccine doses administered in
  Africa</p></title>
  <aug>
    <au><snm>(UNICEF)</snm><fnm>PR</fnm></au>
  </aug>
  <source>UNICEF</source>
  <pubdate>2021</pubdate>
  <url>https://www.unicef.org/press-releases/first-covid-19-covax-vaccine-doses-administered-africa</url>
</bibl>

<bibl id="B103">
  <aug>
    <au><snm>CDC</snm><fnm>A</fnm></au>
  </aug>
  <pubdate>2021</pubdate>
  <url>https://africacdc.org/covid-19-vaccination/</url>
</bibl>

<bibl id="B104">
  <title><p>The benefit of the doubt or doubts over benefits? A systematic
  literature review of perceived risks of vaccines in European
  populations</p></title>
  <aug>
    <au><snm>Karafillakis</snm><fnm>E</fnm></au>
    <au><snm>Larson</snm><fnm>HJ</fnm></au>
    <au><cnm>others</cnm></au>
  </aug>
  <source>Vaccine</source>
  <publisher>Elsevier</publisher>
  <pubdate>2017</pubdate>
  <volume>35</volume>
  <issue>37</issue>
  <fpage>4840</fpage>
  <lpage>-4850</lpage>
</bibl>

<bibl id="B105">
  <title><p>Key lessons from Africa's COVID-19 vaccine rollout</p></title>
  <aug>
    <au><snm>Africa</snm><fnm>WHO</fnm></au>
  </aug>
  <source>World Health Organization</source>
  <publisher>World Health Organization</publisher>
  <pubdate>2021</pubdate>
  <url>https://www.afro.who.int/news/key-lessons-africas-covid-19-vaccine-rollout</url>
</bibl>

<bibl id="B106">
  <title><p>Covid-19 vaccinations: African nations miss who target</p></title>
  <aug>
    <au><snm>Mwai</snm><fnm>P</fnm></au>
  </aug>
  <source>BBC News</source>
  <publisher>BBC</publisher>
  <pubdate>2021</pubdate>
  <url>https://www.bbc.com/news/56100076</url>
</bibl>

<bibl id="B107">
  <title><p>COVID-19 vaccine equity and booster doses</p></title>
  <aug>
    <au><snm>Equity</snm><fnm>V</fnm></au>
  </aug>
  <source>Lancet Infect Dis</source>
  <pubdate>2021</pubdate>
  <volume>21</volume>
  <fpage>743</fpage>
</bibl>

<bibl id="B108">
  <title><p>COVID-19 vaccine hesitancy in Africa: a call to action</p></title>
  <aug>
    <au><snm>Mutombo</snm><fnm>PN</fnm></au>
    <au><snm>Fallah</snm><fnm>MP</fnm></au>
    <au><snm>Munodawafa</snm><fnm>D</fnm></au>
    <au><snm>Kabel</snm><fnm>A</fnm></au>
    <au><snm>Houeto</snm><fnm>D</fnm></au>
    <au><snm>Goronga</snm><fnm>T</fnm></au>
    <au><snm>Mweemba</snm><fnm>O</fnm></au>
    <au><snm>Balance</snm><fnm>G</fnm></au>
    <au><snm>Onya</snm><fnm>H</fnm></au>
    <au><snm>Kamba</snm><fnm>RS</fnm></au>
    <au><cnm>others</cnm></au>
  </aug>
  <source>The Lancet. Global Health</source>
  <publisher>Elsevier</publisher>
  <pubdate>2021</pubdate>
</bibl>

<bibl id="B109">
  <title><p>COVID-19 vaccine hesitancy worldwide: a concise systematic review
  of vaccine acceptance rates</p></title>
  <aug>
    <au><snm>Sallam</snm><fnm>M</fnm></au>
  </aug>
  <source>Vaccines</source>
  <publisher>Multidisciplinary Digital Publishing Institute</publisher>
  <pubdate>2021</pubdate>
  <volume>9</volume>
  <issue>2</issue>
  <fpage>160</fpage>
</bibl>

<bibl id="B110">
  <title><p>The long road ahead for COVID-19 vaccination in Africa</p></title>
  <aug>
    <au><snm>Jerving</snm><fnm>S</fnm></au>
  </aug>
  <source>The Lancet</source>
  <publisher>Elsevier</publisher>
  <pubdate>2021</pubdate>
  <volume>398</volume>
  <issue>10303</issue>
  <fpage>827</fpage>
  <lpage>-828</lpage>
</bibl>

<bibl id="B111">
  <title><p>Eritrea has not started vaccinating against COVID, says Africa
  CDC</p></title>
  <aug>
    <au><cnm>Reuters</cnm></au>
  </aug>
  <source>Reuters</source>
  <publisher>Reuters</publisher>
  <pubdate>2021</pubdate>
  <url>https://www.reuters.com/business/healthcare-pharmaceuticals/eritrea-has-not-started-vaccinating-against-covid-says-africa-cdc-2021-12-09/</url>
</bibl>

<bibl id="B112">
  <title><p>COVID-19 vaccine trials in Africa</p></title>
  <aug>
    <au><snm>Makoni</snm><fnm>M</fnm></au>
  </aug>
  <source>The Lancet Respiratory Medicine</source>
  <publisher>Elsevier</publisher>
  <pubdate>2020</pubdate>
  <volume>8</volume>
  <issue>11</issue>
  <fpage>e79</fpage>
  <lpage>-e80</lpage>
</bibl>

<bibl id="B113">
  <title><p>Booster doses for inactivated COVID-19 vaccines: if, when, and for
  whom</p></title>
  <aug>
    <au><snm>Croda</snm><fnm>J</fnm></au>
    <au><snm>Ranzani</snm><fnm>OT</fnm></au>
  </aug>
  <source>The Lancet Infectious Diseases</source>
  <publisher>Elsevier</publisher>
  <pubdate>2021</pubdate>
</bibl>

<bibl id="B114">
  <title><p>Considerations in boosting COVID-19 vaccine immune
  responses</p></title>
  <aug>
    <au><snm>Krause</snm><fnm>PR</fnm></au>
    <au><snm>Fleming</snm><fnm>TR</fnm></au>
    <au><snm>Peto</snm><fnm>R</fnm></au>
    <au><snm>Longini</snm><fnm>IM</fnm></au>
    <au><snm>Figueroa</snm><fnm>JP</fnm></au>
    <au><snm>Sterne</snm><fnm>JA</fnm></au>
    <au><snm>Cravioto</snm><fnm>A</fnm></au>
    <au><snm>Rees</snm><fnm>H</fnm></au>
    <au><snm>Higgins</snm><fnm>JP</fnm></au>
    <au><snm>Boutron</snm><fnm>I</fnm></au>
    <au><cnm>others</cnm></au>
  </aug>
  <source>The Lancet</source>
  <publisher>Elsevier</publisher>
  <pubdate>2021</pubdate>
  <volume>398</volume>
  <issue>10308</issue>
  <fpage>1377</fpage>
  <lpage>-1380</lpage>
</bibl>

<bibl id="B115">
  <title><p>Israel begins Pfizer booster shots for at-risk adults as delta
  cases rise</p></title>
  <aug>
    <au><snm>Lieber</snm><fnm>D</fnm></au>
  </aug>
  <source>Wall Street journal (Eastern ed.)</source>
  <publisher>wsj.com</publisher>
  <pubdate>2021</pubdate>
  <url>https://www.wsj.com/articles/israel-begins-pfizer-booster-shots-for-at-risk-adults-as-delta-cases-rise-11626110100</url>
</bibl>

<bibl id="B116">
  <title><p>COVID-19 vaccine booster shots</p></title>
  <aug>
    <au><cnm>CDC</cnm></au>
  </aug>
  <source>Centers for Disease Control and Prevention</source>
  <pubdate>2022</pubdate>
  <url>https://www.cdc.gov/coronavirus/2019-ncov/vaccines/booster-shot.html</url>
</bibl>

<bibl id="B117">
  <title><p>WHO director general's opening remarks at the media briefing on
  COVID 4 August 2021</p></title>
  <source>Who.int</source>
  <publisher>WHO-Director General</publisher>
  <pubdate>2021</pubdate>
  <url>https://www.who.int/director-general/speeches/detail/who-director-general-s-opening-remarks-at-the-media-briefing-on-covid-4-august-2021</url>
</bibl>

</refgrp>
} 


\end{backmatter}
\end{document}